%Paper: gr-qc/9403036
%From: avramidi@math-inf.uni-greifswald.d400.de
%Date: Fri, 18 Mar 1994 19:31:18 +0100

%
% Paper: gr-qc/9403036
% plain TeX (EMTeX)
% From:  "I. G. Avramidi" <avramidi@mathematik.uni-geifswald.d400.de>
% Date: 18 March, 1994
%
%  submitted to:  ANNALS OF PHYSICS
%

\tolerance=2000
\hbadness=2000
\overfullrule=0pt
\magnification=1200
\baselineskip=12pt

\font\gross=cmbx10 scaled \magstep2

\font\sc=cmcsc10
\font\bfm=cmmib10

\def\RR{\rm I\!R}
\def\h#1{{\cal #1}}
\def\a{\alpha}
\def\abf{{\bfm \char'013}}
\def\b{\beta}
\def\g{\gamma}
\def\d{\delta}
\def\eps{\varepsilon}
\def\l{\lambda}

\def\m{\mu}
\def\mbf{{\bfm \char'026}}
\def\n{\nu}
\def\nbf{{\bfm \char'027}}
\def\s{\sigma}

\def\r{\rho}
\def\na{\nabla}

\def\square#1{\mathop{\mkern0.5\thinmuskip\vbox{\hrule\hbox
	{\vrule\hskip#1\vrule height#1 width 0pt\vrule}\hrule}
         \mkern0.5\thinmuskip}}
\def\Square{\mathchoice
	{\square{6pt}}{\square{5pt}}{\square{4pt}}{\square{3pt}}}
\def\sq{\Square}

%
% end of definitions
%

{\nopagenumbers
\pageno=1
\null

\centerline{\gross COVARIANT  \ METHODS  \ FOR  \ CALCULATING}
\smallskip
\centerline{\gross THE  \ LOW-ENERGY \ EFFECTIVE \  ACTION}
\smallskip
\centerline{\gross IN  \ QUANTUM \  FIELD \  THEORY }
\smallskip
\centerline{\gross  AND \  QUANTUM \ GRAVITY}

\bigskip
\centerline{I. G. Avramidi
\footnote{\dag}{Alexander von Humboldt Fellow}
\footnote{\S}{This work was supported, in part by a Soros Humanitarian
Foundation's
 Grant awarded by the American Physical Society and by an Award
  through the International Science Foundation's Emergency Grant
  competition.}
\footnote{\ddag}{On leave of absence from Research Institute
 for Physics, Rostov State University,  Stachki 194, 344104 Rostov-on-Don,
 Russia.} }
\bigskip
\centerline{\it
Department of Mathematics, University of Greifswald}
\centerline{\it
Friedrich-Ludwig-Jahn-Str. 15A, 17489 Greifswald, Germany.}

\bigskip
\bigskip
\centerline{\sc Abstract}
\bigskip
We continue the development of the effective covariant methods for
calculating the heat kernel and the one-loop effective action in quantum
field theory and quantum gravity. The status of the  low-energy approximation
in quantum gauge theories and quantum gravity is discussed in detail on the
basis of analyzing the local Schwinger - De Witt expansion. It is argued that
the low-energy limit, when defined in a covariant way, should be related
to background fields with covariantly constant curvature, gauge field
strength and potential.  Some new approaches for calculating the low-energy
heat kernel assuming a covariantly constant background are proposed.  The
one-loop low-energy effective action in Yang-Mills theory in flat space with
arbitrary compact simple gauge group and arbitrary matter on a covariantly
constant background is calculated.  The stability problem of the
chromomagnetic (Savvidy-type) vacuum is analyzed. It is shown, that this type
of vacuum structure can be stable only in the case when more than one
background chromomagnetic fields are present and the values of these fields
differ not greatly from each other.This is possible only in space-times of
dimension not less than five $d\geq 5$.
\medskip

PACS number(s): 03.70.+k, 04.60.+n, 04.90.+e, 02.40.Vh, 12.38.Aw-t
\vfill
\eject}

\centerline{\bf I. INTRODUCTION}
\bigskip

In present paper we continue our efforts in developing methods for computing
the effective action in quantum field theory, that we started in
[1-7].
The effective action is a very powerful tool for investigating the general
problems of quantum field theory as well as various models. It takes into
account all quantum fluctuations and contains, in principle, all predictions
of quantum field theory
[8-13].
Due to special advantages achieved by using geometric methods
the effective action approach
 is spread widely in gauge theories and quantum gravity, supergravity,
Kaluza-Klein models, strings etc.
[14-20].

However, the practical calculation of the effective action will entail great
difficulties. The point is, that, although it is possible to calculate the
effective action in some rare special cases on fixed background, for
effective action to be used one has to vary it, i.e. one needs  it
as a functional of background fields of general type.
Therefore, various approximate methods for calculating  the effective action
were elaborated. The first one is the so called Schwinger - De Witt expansion
[8,9-13,21,22],
which was successfully used for treating divergences, renormalization,
anomalies etc.
[20,17].

All quantities of interest (such as the effective action,
Green function, stress tensor, currents, anomalies) are expressed in this
approach in terms of coefficients of asymptotic expansion of the
corresponding heat kernel, so called Hadamard - Minakshisundaram - De Witt -
Seely (HMDS) coefficients
[23-44].
Various methods were used for calculating these coefficients, beginning
from the direct De Witt's method
[9,10]
to modern mathematical methods, which make use of pseudodifferential operators
[28,32,42].
Very good review of calculation of HMDS-coefficients is given in recent paper
[34].
The first three coefficients were calculated in
[28].

An effective covariant technique for calculating HMDS-coefficients is
elaborated in
[1-6],
where also the first four coefficients are calculated. In the case of scalar
operators the fourth coefficient is also calculated in
[45].
About the fifth coefficient in flat space see
[46].
Analytic approach was developed in
[38,39],
where a general expression in closed form for these coefficients was
obtained.

The Schwinger - De Witt expansion describes good the vacuum polarization
effect of massive fields in weak background fields
[8,10,20,47].
But it is absolutely inadequate and becomes meaningless in strong background
fields and massless theories. For investigating these (nonlocal) aspects of
the effective action one needs some new methods. There were proposed,
 on the one hand, the immediate direct partial summation of higher
derivatives in local Schwinger - De Witt expansion
[13,2,3,5,6], and, on the other hand,
covariant perturbation theory  in powers of curvatures
[48,49].
Both of these methods lead finally to nonlocal expansion of the effective
action, which is valid also in massless theories but only in high-energy
 (short-wave) limit, when the background curvatures are  small rapidly
varying external fields. This approximation was used already for calculating
the anomalous magnetic moment of the electron
[50]
and investigating the Hawking radiation in two dimensions and the
gravitational collapse problem
[13,51,52].

Methods and results of the papers
[48,49],
are valid, strictly speaking, for asymptotically flat noncompact
complete manifolds without the boundary. They can not help in
analyzing long-wave global aspects of the effective action. The low-energy
effective action (in other words, the effective potential) presents a very
natural tool for investigating the vacuum of the
theory, its stability and the phase structure
[14-18].
It is determined by strong slowly varying background fields and, therefore,
its calculation depends essentially on global topological properties of the
space-time manifold.  This is very important and, in general case, still
unsolved problem. Some new results on this subject concerning calculation
of the low-energy heat kernel and one-loop effective potential are  obtained
in [67, 68].

All these approximations are bound up with each other. They are, as a matter
of fact, various reexpansions of a single general effective action of quantum
gravity.

This  paper is organized as follows. The heat kernel method and the Schwinger
- De Witt expansion is described in the next sect. II. We analyze here also
the general structure of the HMDS - coefficients. In sect. III we discuss the
status of the high-energy approximation and the low-energy one in quantum
gravity and gauge theories.  It is found that a symmetry Lie group
 appears when formulating the low-energy approximation in
covariant way, i.e on covariantly constant background.  A number of
perspective methods for calculating the effective action in low-energy
approximation, in which, in particular, the symmetry algebra, is used
essentially, are proposed in sect.  IY.  Sect.  Y is devoted to the
calculation of the low-energy effective action in flat space.  Here a Yang -
Mills model with any matter fields is considered, the corresponding one-loop
effective potential is computed and the vacuum structure of the model is
slightly analyzed. In the concluding sect. YI we formulate briefly the basic
results.

We use in this paper Euclidean notations meaning to obtain physical
effective equations by analytic continuation. As it is shown in
[48]
this can be done for a very wide class of problems.
 Our notations and definitions of curvature tensor are as in
[9,12].

\bigskip
\bigskip
\centerline{\bf  II. ONE-LOOP EFFECTIVE ACTION AND HEAT KERNEL}
\bigskip

For a wide range of models in quantum field theory the contribution of a
multiplet of quantized fields $\varphi=\{\varphi^A(x)\}$  on $d$-dimensional
Riemannian manifold $M$  to the effective action $\Gamma$ in one-loop
approximation is expressible in terms of a functional superdeterminant
of an elliptic  second order differential operator
$$
\eqalignno{
&\Gamma = S+\hbar \Gamma_{(1)} + {\rm O}(\hbar^2)&(2.1)\cr
&\Gamma_{(1)}={1\over 2}\ln{\rm Sdet}{F\over \m^2} =
{1\over 2}{\rm Str}\ln{F\over \m^2}                   &(2.2)\cr
&F=g^{1/2}(-\Square + Q + m^2)                        &(2.3)\cr
&\sq=g^{\m\n}\na_\m\na_\n=g^{-1/2}
(\partial_\m+\h A_\m)g^{1/2}g^{\m\n}(\partial_\n+\h A_\n)&(2.4)\cr}
$$
where $S$ is a classical action of the model,
$Str$ means functional supertrace, $\m$ is a renormparameter,
$g_{\m\n}$ is the metric, $g=\det g_{\m\n}\ne 0$ , $Q= \{ Q^A_{\
B}(x)\}$ is an arbitrary matrix (potential term), $m$ is a mass parameter and
$\na_\m$ is a covariant derivative, defined with the help of an
arbitrary linear connection $\h A_\m$, including both vector gauge
and appropriate spin one. As regards the more general differential operators
see
[21].

Let us mention that the multiplet  $\varphi=\{\varphi^A\}$ realizes, in
general, reducible representation of gauge and Lorentz group. That means,
that it contains all physical fields, i.e. scalar, spinor, vector etc. ones
 $\varphi=\{\varphi^a(x),\psi^i(x),\h B^c_\m,h_{\m\n},...\}$.
We assume also operator  $F$ (2.2) to be positive. This is, in fact, the case
when $m^2$ is sufficiently large and the background fields are regular
and bounded at the infinity.

Thus the effective action is a functional of the coefficient functions of the
operator $F$ (2.2), i.e. three different background fields - the metric,
the connection and the potential term $\Gamma=\Gamma(g_{\m\n},\h
A_\m,Q)$.

Making use of the covariance of the operator  $F$ (2.2),
it is easy to show, that the effective action (2.1) possess a crucial
property - it is invariant under diffeomorphisms and local gauge
transformations.
That means that the effective action depends, in fact, only on the geometry of
the manifold and on the classes of gauge equivalent fields and does not
depend on the choice of the set of coordinates and the gauge of gauge
field.

Local invariant characteristics of the background fields are described
by covariant quantities (tensors),i.e. Riemann tensor $R_{\m\n\a\b}$
and curvature of background connection
$$
\h R_{\m\n} = \partial_\m\h A_\n-\partial_\n\h A_\m+[\h A_\m,\h A_\n]
\eqno(2.5)
$$
(along with the potential term $Q$, which also transforms covariantly).
These three quantities have the same dimensions and play often the same
role in calculations. That is why we shall call them generalized
background curvatures (or simply curvatures) and denote in symbolic
expressions by one symbol
$$
R=\{R_{\m\n\a\b}, R_{\m\n}, Q\}
$$

The effective action is, in general, very complicated nonlocal functional and
must be built, therefore, from covariant geometric objects. Those can be
either local tensors (such as curvature tensors and their derivatives) or
nonlocal (two-point, three-point etc.) covariant geometric objects, which
transform completely independent as different tensors in different points of
the manifold.  The most commonly encountered nonlocal covariant  geometric
objects are the world function (or the geodetic interval), the parallel
displacement operator and Green functions of some covariant differential
operators.

The geometry of the manifold is described, in general, not only by local
characteristics like invariants of the curvature tensor, but also by global
structure - its topology, boundaries etc. The curvature tensor alone does not
determine all details of the geometry of the whole manifold.

The effective action is determined, generally speaking, by the spectrum of
the operator $F$ (2.2) and, naturally, depends also on the global structure
of the manifold. Thus the effective action depends, in general,
also on the topology. Let us mention, that this does not contradict with the
covariance, what is meant sometimes when one calls the contributions
in the effective action, which can not be expressed locally in terms of the
curvature, noncovariant.  The covariance is, simply, a local concept, and
the topology is a global one.  Therefore one can speak about the covariance
of the effective action within the limits of fixed topology, as the
diffeomorphisms and local gauge transformations do not lead out of bounds of
the given topology.

Let us  mention from the very beginning, that in this paper we will be
interested in the case of complete noncompact asymptotically flat manifolds
without boundaries with standard boundary conditions, viz., the regularity
on the Euclidean infinity. We will not investigate  the influence
of the topology and will concentrate our attention, as a rule,  on the
effects of the curvature.

For covariant calculation of the effective action it is convenient to
express it in terms of the heat kernel $U(t)$
$$
\eqalignno{
&\Gamma_{(1)}=\int dx \left(-{1\over 2}\right)
\int\limits^\infty_0{d\,t\over t}{\rm str}U(t|x,x) &(2.6)\cr
&U(t|x,x') = \exp(-tH)\h P(x,x')\d(x,x')                 &(2.7)\cr
&H = g^{-1/4}Fg^{-1/4}
= -\sq + Q + m^2
&(2.8)\cr}
$$
 where $str$ is a usual matrix supertrace and  $\h P(x,x')=\{\h P^A_{\
 B'}(x,x')\}$ is the parallel displacement operator of the field
 $\varphi=\{\varphi^A\}$  from the point
 $x$  to the point $x'$  along the geodesic,
[53,9,12,6].

Thus the main object to deal with is the heat kernel $U(t)$ (2.7).
More explicitly it is defined by requiring it to satisfy the equation
$$
\left({\partial\over\partial t}+H\right)U(t\vert x,x^\prime)=0
\eqno (2.9)
$$
with initial condition
$$
U(0\vert x,x')=\d(x,x')
 \eqno (2.10)
$$
In the case of noncomplete manifolds one should also impose some boundary
conditions
[54-59].
However we will not consider in this paper the boundary effects.

It is well known that the heat kernel at small $t\to 0$ and when the points
 $x$ and $x'$ are close to each other $x\to x'$ behaves like
[9,10,31,]
$$
\eqalignno{
U(t\vert x,x') = &g^{1/4}(x)g^{1/4}(x')\Delta^{1/2}(x,x')\h P(x,x')
\exp\left(-{\s(x,x')\over 2t}\right)&\cr
&\times(4\pi t)^{-d/2}\exp\left(-tm^2\right)
\sum\limits_{k=0}^\infty {(-t)^k\over k!} a_k(x,x')&(2.11)\cr}
 $$
where $\s(x,x')$ and $\Delta(x,x')$  are the geodetic interval (world
function), which is equal to one half the square of the distance along the
geodesic between the points $x$ and $x'$, and the corresponding Van Vleck -
Morette determinant
[53,62,9,12].

This expansion at coinciding points $x=x'$ is called often the Schwinger -
De Witt one. We call its coefficients $a_k(x,x)$  the Hadamard -
Minakshisundaram - De Witt - Seely (HMDS) coefficients. One can show, that
Schwinger - De Witt expansion is purely local and HMDS-coefficients are local
invariants built from the curvatures (including the potential term) and their
covariant derivatives
[9,10,12,31,6].
       They play very important role in physics as well as in mathematics
       and are closely connected with various sections of mathematical
       physics
[34,31].
Therefore, the calculation of HMDS-coefficients is in itself of great
importance. Nowadays in general case only the first four coefficients are
calculated
[1,4,5,6,45].
        In manifolds with boundary additional terms in the asymptotic
       expansion of the heat kernel proportional to $t^{-d/2+k/2}$ appear.
For details see
[54-59,7].

Several first HMDS-coefficients have the form
        $$
        \eqalignno{
	a_0(x,x) &=1            &\cr
        a_1(x,x) &=Q-{{1}\over {6}} R      &(2.12)\cr
  a_2(x,x) &=\left(Q-{{1}\over {6}} R\right)^2-{1\over 3}\sq Q
  +{1\over 6}\h R_{\m\n}\h R^{\m\n }-{1\over 90}R_{\m\n}R^{\m\n}
  + {{1}\over {90}} R_{\m\n\a\b}R^{\m\n\a\b}+{{1}\over {15}}\sq R
        &\cr}
       $$
       The general structure of other  coefficients  $a_k$ can be represented
       symbolically in the form
$$
\eqalignno{
a_k(x,x) &= \na^{2k-2}R +
       \sum_{{ 0 \le i \le 2k-4}}\na^i R\ \na^{2k-4-i} R +
       \sum_{{i,j \ge 0;\ i+j\le 2k-6}}
       \na^i R\ \na^j R\  \na^{2k-6-i-j}R &\cr
       &+ \cdots +
       \sum_{0 \le i \le k-1}R^i(\na\na R) R^{k-i-1} +
       \sum_{{i,j \ge 0;\ i+j\le k-3}}R^i(\na R)R^j(\na R) R^{k-i-j-3} +
       R^k &\cr
       & &(2.13)\cr}
$$
       From the behavior of the heat kernel at $t\to 0$ (2.11) it is
       immediately seen, that the expression (2.6) for the effective action
       diverges at the lower limit of integration  $t\to 0$.
Therefore, one has to regularize and to renormalize it, the
covariance being manifestly preserved. The most convenient way to do it in
the one-loop case is  to use the  $\zeta$-function prescription
$$
\eqalignno{
&\Gamma_{(1)}=-{1\over 2}\zeta'(0)           &(2.14)\cr
&\zeta'(0)={d\over dp}\zeta(p)\big\vert_{p=0}&(2.15)\cr}
$$
where
 $$
\zeta(p)=\m^{2p}{\rm Str}H^{-p}=\int dx
{\m^{2p}\over \Gamma (p)}\int\limits_0^\infty dt\
t^{p-1}{\rm str}U(t|x,x)
 \eqno (2.16)
 $$

The main merit of the heat kernel method is that the heat kernel is good
defined and does not have any divergences in contrast to the effective action.
This allows to single out in the effective action the property of the
covariance from the field-theoretical divergences problem. In  other words,
making use of the heat kernel divides the problem of evaluating the effective
action in two parts: {\sl i)} calculating the heat kernel and {\sl ii)}
regularization and renormalization.

Using the $\zeta$-function regularization (2.14) and the Schwinger - De Witt
expansion (2.11) one can get the asymptotic $1/m^2$-expansion for the
effective action in terms of HMDS-coefficients. One obtains in this way
 for odd dimension
$$
\Gamma _{(1)}={1\over 2}(4\pi )^{-d/2}
\pi (-1)^{d-1\over 2}
\sum\limits_{k\ge 0}{m^{d-2k}\over k!\Gamma \left({1\over 2}d-k+1\right)}A_k
\eqno(2.17)
$$
and for even dimension
$$
\eqalignno{
\Gamma_{(1)}={1\over 2}(4\pi )^{-d/2}\Biggl\{&(-1)^{d/2}
\sum\limits _{k=0}^{d/2}
{m^{d-2k}\over k!\Gamma \left({1\over 2}d+1-k\right)}
A_k\left[\ln {m^2\over \mu^2}-
\Psi \left({d\over 2}-k+1\right) -{\bf C} \right]&\cr
&+\sum\limits_{k\ge {1\over 2}d+1}
{\Gamma \left(k-{d\over 2}\right)(-1)^k\over k!m^{2k-d}}A_k
\Biggr\}&(2.18)\cr}
$$
where
$$
A_k=\int d x g^{1/2}{\rm str}\,a_k(x,x)
\eqno(2.19)
$$
 $$
 \Psi (q)={d\over dq}\ln
 \Gamma (q)\ \ \ , \ \ \ {\bf C}=-\Psi (1)
 $$
 Let us notice the essentially different dependence of the effective action
 on the renormparameter $\m$  in the spaces of odd and even dimensions.
 This is due to finiteness of one-loop effective action in odd dimensions.

 Thus the Schwinger - De Witt expansion of the effective action
 (2.17), (2.18) is purely local and does not depend, in fact, on the
 global structure of the manifold. Using the integration by parts one can
 easily conclude that the general structure of the coefficients $A_k$
 (2.19) has the form
$$
\eqalignno{
A_k &= \int dx g^{1/2}{\rm str}\,\Biggl\{
        R \sq^{k-2} R +
       \sum_{ 0\le i \le 2k-6} R\ \na^i R\  \na^{2k-6-i}R &\cr
       &+ \cdots + \sum_{0\le i\le k-3}R^i(\na R)R^{k-i-3}(\na R)  +
       R^k \Biggr\}
       &(2.20)\cr}
$$

It is evident, that the local Schwinger - De Witt expansion is a good
approximation and describes, therefore, the vacuum polarization effect of
massive quantum fields in weak background fields when
 $$
 A_k \ll m^{2k}
 $$
i.e. all invariants of the curvature are smaller than the corresponding power
of the mass parameter. However it is not good approximation  in the case of
strong background fields and absolutely meaningless in massless theories. For
investigation of these cases one needs other special methods, which should
also be manifestly covariant.

One possibility to exceed the limits of the Schwinger - De Witt
expansion in manifestly covariant manner is to compare all the terms in
HMDS-coefficients (2.20), to pick up the main (the largest in some
approximation) terms and to sum up the corresponding partial sum. This idea
was first proposed in
[13]
and realized in
[2,3,5,6],
where all the terms in coefficients $A_k$ (2.20) with higher derivatives,
 which are quadratic in curvatures are calculated and the Schwinger - De Witt
 (2.11) expansion is summed in this approximation. The terms quadratic in
 curvatures were also calculated completely independent in
 [60].
 In recent papers
 [22,15,51]
   the third order in curvatures is analyzed.

\bigskip
\bigskip
\centerline{\bf III. APPROXIMATIONS UNDER CONSIDERATION}
 \bigskip

%  begin  of definitions  %

\def\nohkf {+ {{t^2}\over {2}}\biggl[Q\gamma^{(1)}(t\Square)Q+ 2\h
            R_{\alpha\mu}\nabla^\alpha{{1}\over {\Square}}\gamma^{(2)}
             (t\Square)\nabla_\nu\h R^{\nu\mu} - 2Q\gamma^{(3)}(t\Square)R}
\def\nohks {+R_{\mu\nu}\gamma^{(4)}(t\Square)R^{\mu\nu}
              +R\gamma^{(5)}(t\Square)R\biggr] }

%  end of definitions  %

Thus the effective action is a functional of the metric, connection and the
potential term. This means that it depends on all  details of the behavior
of these functions of the coordinates.  Exact evaluation of the effective
action is impossible even in one-loop approximation.  Therefore, one should
use some approximations based on the assumptions about the arguments of that
 functional, i.e. about the behavior of these quantities as functions of
 coordinates.

A remark concerning the meaning of the background fields as functions of
coordinates has to be made here. Namely, one should distinguish the
real physical variations of these functions from the purely gauge fictive
ones, which reflect simply the transformations of the coordinates and local
gauge transformations. In other words, one has to factorize out the gauge
degrees of freedom of background fields. It is  this physical gauge invariant
dependence on the coordinates that we will mean when speaking about the
behavior of the background fields.  This can be achieved by using  for local
 description of background fields the invariants (scalars) built from the
 curvatures and by comparing the values of invariants themselves with  the
derivatives of them.  Since the usual derivatives of the invariants coincide
with the covariant derivatives, one can compare the values of curvatures and
their covariant derivatives.

If one expands the background fields in the Fourier integral, which can be
defined in an appropriate way, then one may regard the effective action to
depend on the Fourier components of background fields. Here  two main
approximations are possible.

\bigskip
\centerline{\sc 3.1. High-energy approximation}
 \medskip

First of all, this is the high-energy limit, when the details of the
dependence of the effective action on the short-wave part of the spectrum of
the background fields are analyzed. In coordinate representation this
approximation corresponds to weak rapidly varying background fields
(ripples) that may be characterized by a covariant relation
$$
\na\na R \gg RR
\eqno(3.1)
$$
Since the commutators of the covariant derivatives add a spare power of the
curvature, they commute at a given order in this approximation.

This approximation corresponds to partial summation at first in local
Schwinger - De Witt expansion (2.11)  the terms with higher derivatives (in
 other words with higher momentums) of background fields. Analyzing and
 summing up such terms
[2,3,5,6]
leads to explicit covariant nonlocal expressions for the heat kernel
     $$
     \eqalignno{
  {\rm Str}\,U(t) &=
     \int dx\, g^{1/2}(4\pi t)^{-d/2}\exp{(-tm^2)}str\biggl\{
                1-t\left(Q-{{1}\over {6}}R\right)               & \cr
                &\nohkf                                    &\cr
                &\nohks             +O(R^3)\biggr\}\qquad . &(3.2)\cr}
     $$
and the effective action
      $$
     \Gamma_{(1)}=\Gamma_{(1)loc}+\Gamma_{(1)nonloc}      \qquad .
\eqno(3.3)
     $$
     where the local part equals: in odd dimension
     $$
     \Gamma_{(1)loc} ={{1}\over {2}}(4\pi)^{-d/2}{{\pi (-1)^{{{d-1}
                       \over {2}}
                        }}\over {\Gamma({{d}\over {2}} +1)}}
                        \int dx\,g^{1/2}str\left\{m^d
                        +{{d}\over {2}}m^{d-2}\left(Q-{{1}\over {6}}R\right)+
                     O(R^3)\right\}
\eqno(3.4)
     $$
and in even dimension
     $$
     \eqalignno{
     \Gamma_{(1)loc} &={{1}\over {2}}(4\pi)^{-d/2}{{(-1)^{d/2}}\over {\Gamma
                        ({{d}\over {2}}+1) }}\int dx\,g^{1/2}str\biggl\{
                        m^d\left[\ln {{m^2}\over {\mu^2}}
                        -\Psi\left({{d}\over {2}}\right)
                        -{\bf C}\right]                               & \cr
                       &+{{d}\over {2}}m^{d-2}\left[\ln{{m^2}\over {\mu^2}}
                       -\Psi\left({{d}\over {2}}\right)-{\bf C}\right]\left(Q
                       -{{1}\over {6}}R\right)
                       +O(R^3)\biggr\}     \qquad .           &(3.5)\cr}
      $$
       \par
       \noindent
       The nonlocal part of the effective action has the form
       $$
       \eqalignno{
       \Gamma_{(1)nonloc}&={1\over 2}(4\pi )^{-d/2}\int dx \,g^{1/2}\,str
                           \biggl\{Q\beta^{(1)}(\Square)Q
                           +2\h R_{\alpha\mu}\nabla^\alpha{{1}\over {\Square}}
                           \beta^{(2)}(\Square)\nabla_\nu\h R^{\nu\mu}& \cr
                          &-2Q\beta^{(3)}(\Square)R
                          +R_{\mu\nu}\beta^{(4)}(\Square)R^{\mu\nu}
                          +R\beta^{(5)}(\Square)R
                          +O(R^3)\biggr\}                   &(3.6)\cr}
        $$
Here $\g^{(i)}(t\sq)$ and $\b^{(i)}(\sq)$  are formfactors, which are obtained
explicitly in general case in
[2,3,5,6].
In papers
[48,49]
a similar expressions for massless case $m=0$ was obtained using completely
different method, viz. the covariant perturbation theory proposed
in these papers. Next third order in curvatures is under investigation in
[15,22,51].

\bigskip
\centerline{\sc 3.2. Low-energy approximation}
 \medskip

 Let us now discuss the opposite approximation, i.e. the long-wave (or
 low-energy) one. It can be characterized by an opposite relation
$$
\na\na R \ll RR
\eqno(3.7)
$$
This means that the derivatives of all invariants of background fields are
much smaller than the products of the invariants themselves.

In this case the trace of the heat kernel and the effective action should
have expansions of the form
$$
     \eqalignno{
    &{\rm Str}\,U(t) =
     \int dx\, g^{1/2}(4\pi t)^{-d/2}\exp{(-tm^2)}{\rm str}\left\{
     \Phi(tR) + \na R\Psi(tR)\na R + {\rm O}(\na^3) \right\}& \cr
&\Gamma_{(1)}= \int dx\, g^{1/2}{\rm str}\left\{V(R) + \na R Z(R)\na R +
{\rm O}(\na^3) \right\}& (3.8)\cr}
$$
where $\Phi(tR), \Psi(tR), V(R)$ and $Z(R)$ are local functions and
symbol $ O(\na^3)$ means the terms with more than three derivatives.

It is naturally to call the first term in the formula for
the effective action (3.8) (the zero order),  which does not contain the
covariant derivatives at all, the effective potential of quantum
gravity
$$
\Gamma_{(1)}\vert_{\na R=0}= \int dx\, g^{1/2}{\rm str}\,V(R)
\eqno(3.9)
$$

Let us mention that such a definition of the effective potential and the
expansion in covariant derivatives is not conventional. It differs from
the definition that is often found in the literature
[19,44].
What is meant usually under the notion of the effective potential is a
function of the potential term only $Q$, because it does not  contain
derivatives of the background fields and transforms by itself  covariantly
(in contrast to curvatures that contain first $\h R \sim \partial \h A$
and second $R_{\m\n\a\b}\sim \partial\partial g$ derivatives of the
background fields).  So, e.g. in
[19,44]
the  potential term $Q$ is summed up exactly but an expansion is made not
only in covariant derivatives but also in powers of curvatures $R_{\m\n\a\b}$
and $\h R_{\m\n}$, i.e. the curvatures are treated perturbatively, since they
contain the derivatives of the fields. Thereby the validity of this
approximation for the effective action  is limited to  small curvatures
$$
R_{\m\n\a\b}, \h R_{\m\n} \ll Q
\eqno(3.10)
$$
Such an expansion is called `expansion of the effective action in covariant
derivatives'. Without the potential term   $Q=0$ the effective potential in
such a scheme is trivial. Hence we stress here once again, that in our
definition the effective potential contains, in fact,  much more information
than the usual effective potential what is often  meant in the
literature, when using the `expansion in covariant derivatives'. As a matter
of fact, what we mean is the low-energy limit of the effective action
formulated in a gauge invariant way.

It is more difficult to formulate the low-energy approximation in an
invariant way. The point is that in short-wave approximation (3.1), i.e. by
expanding in curvatures, the zero approximation $R=0$ corresponds simply
to the flat empty space, i.e. the standard vacuum of the field theory.
Therefore, the covariant perturbation theory
[48,49]
is, in fact, only the covariant reformulation of the usual noncovariant
perturbation theory. In this case the effective action in a given order can
be built, in principle,  from  several first usual perturbative $n$-point
Green functions. This is reflected by the fact that after arranging the
covariant derivatives in first, second etc. orders in some manner one can fix
uniquely the structure of the whole series (for details see
[13,5,6]).

Nothing of the kind is the relation (3.7), i.e.  the condition of  weak
dependence of the invariant characteristics of the background fields and the
manifold itself on the coordinates.  There is a lack of uniqueness in the
structure of the series for the low-energy effective action. The disposition
of the covariant derivatives in higher orders has an influence on  the form
of the zero approximation. This is related to the fact, that some
combinations of the curvatures (viz. those, which give the commutators of
covariant derivatives) can be treated as terms with two covariant
derivatives. Therefore one has to come to an understanding about the
disposition of the covariant derivatives and to fix the structure of the
series.

One can say that the evaluation of the effective potential (3.9) corresponds
in usual perturbative field theory to summing up all the diagrams at zero
momentums, the calculation of the next term in (3.8) (first order) - to
summing up second derivatives of all diagrams with respect to momentums at
zero momentums etc.. In other words, the calculation of the effective
potential corresponds to the calculation of the  $\infty$-point Green
function in field theory.

The problem is now that the zero approximation is no longer the
empty flat space. The manifold must be strongly curved even in zero
approximation. Hence the various topologies of the background manifold are
possible and one has, in general, to analyze the influence of the topology.

Thus we realize that one should single out in the curvatures the covariantly
constant background $\tilde R$ and the rest part $P$ treat as a
short-wave perturbation (ripple)
$$
\eqalignno{
&R = \tilde R + P&\cr
\tilde\na \tilde R = 0 \qquad,&\qquad
PP \ll \tilde\na\tilde\na P \ll \tilde R\tilde R&(3.11)\cr}
$$
Apparently, one can extend the covariant perturbation theory developed in
[48,49],
to this case, i.e. when the curvature in zero approximation does not
vanish and is covariantly constant. We shall not do this in present paper,
intending to calculate only the zero order, i.e. the effective potential.
So we will not be concerned with this problem.

  In more detail, the zero approximation corresponds to covariantly
constant curvatures
$$
\na_\m R_{\a\b\g\d} = 0 \qquad,\qquad
\na_\m \h R_{\a\b}  = 0 \qquad,\qquad
\na_\m Q = 0
\eqno(3.12)
$$
The conditions of integrability of these relations lead to  strong
algebraic restrictions on the curvatures themselves
$$
\eqalignno{
&R_{\m\n\l[\a}R^\l_{\ \b]\g\d} + R_{\m\n\l[\g}R^\l_{\ \d]\a\b} = 0&(3.13)\cr
&R_{\m\n\l[\a}\h R^\l_{\ \b]} + R_{\a\b\l[\m}\h R^\l_{\ \n]} = 0&(3.13a)\cr
&[\h R_{\m\n},\h R_{\a\b}] + R_{\m\n\l[\a}\h R^\l_{\ \b]}
- R_{\a\b\l[\m}\h R^\l_{\ \n]} = 0&(3.14)\cr
&[\h R_{\m\n},Q] = 0&(3.15)\cr}
$$

Mention that the relation (3.13) is local. It determines locally the geometry
of the symmetric spaces
[61].
However, the manifold is globally symmetric one only in the case when
it satisfies additionally some global topological restrictions and
the condition (3.13) is valid everywhere, i.e. in any point of the manifold
[61].

But in our case, i.e. in physical problems, the situation is radically
different. The correct setting of the problem is as follows. Consider
an asymptotically flat space that is homeomorphic to $\RR^d$  (as we already
stressed above, we do not consider the topological effects, meaning to
investigate, in general, the influence of the curvature).  Let  a finite  not
small, in general, domain of the manifold exists that is strongly curved
and quasi-homogeneous, i.e. the invariants of the curvature in this
region vary very slowly.  Then the geometry of this region is  locally very
similar to that of a symmetric space.  However one should have in mind that
there are always regions in the manifold where this condition is not
fulfilled.  This is, first of all, the asymptotic Euclidean region that has
small curvature and, therefore, the opposite short-wave approximation is
valid.

 Thus the general situation in correct setting of the problem is the
following.  From infinity with small curvature and possible radiation where
[15,22,48,49]
$$
RR\sim |x|^{-2d} \ll \na\na R\sim |x|^{-d-2}
\eqno(3.16)
$$
we pass on to quasihomogeneous region where the local properties of the
manifold are close to those of symmetric spaces. The size of this region can
tend to zero. Then the curvature is nowhere large and the short-wave
approximation is valid anywhere.

If one tries to extend the limits of such region to infinity, then
one has also to analyze the topological properties. The space can be compact
or noncompact depending on the sign of the  curvature. It can be also a
combination of these cases, i.e. a product of a compact and noncompact spaces.
But first we will come across a coordinate horizon-like singularity,
although no one true physical singularity really exist.

This construction can be  intuitively imagined as follows. Take the flat
Euclidean space $\RR^d$, cut out from it a region $M$ with some boundary
and stick to it along the boundary instead of the piece cut  out a piece of a
curved symmetric space with the same boundary $\partial M$.  Such a
construction will be homeomorphic to the initial space and at the same time
will contain a finite highly curved homogeneous region. By the way, if one
subtracts from the exact effective action for a symmetric space the effective
action for built construction then one gets purely topological contribution
to the effective action. This fact seems to be useful when analyzing the
effects of topology.

Thus the problem is to calculate the effective action for covariantly
constant background, i.e. the low-energy effective action. Although this
quantity, generally speaking, depends essentially on the topology and other
global aspects of the manifold, one can disengage oneself  from these effects
fixing the trivial topology. Since the asymptotic Schwinger - De Witt
expansion does not depend on the topology, one can hold that we thereby
sum up all the terms without covariant derivatives in it. In the next
section we adduce several approaches for calculating the effective action.

Let us mention also some ideas and formulae that we will need further for
calculation of the effective potential.

One can use the conditions of covariant constancy of the curvatures (3.12) to
express the metric  and the connection directly in terms of curvatures, i.e.
to integrate explicitly these equations. This can be done in covariant way by
using instead of local quantities like the metric and the connection, some
auxiliary two-point geometric objects - world function (or geodetic interval)
$\s(x,x')$ and parallel displacement operator $\h P(x,x')$.

One can show that the geodetic interval characterizes the geometry of
the manifold in full, i.e. it gives an alternative description of a curved
space
[53,62,9]
and the parallel
displacement operator describes completely the connection.

As it is shown in
[1-6]
one can organize the whole calculating process in such a way that all
quantities will be expressed in terms of derivatives of the following
two-point functions
$$
\eqalignno{
&\s^{\m'}_{\ \n}=\na_\n\na^{\m'}\s            &(3.17)\cr
&X^{\m'\n'} = \s^{\m'}_{\ \a}\s^{\n'\a}&(3.17a)\cr
&\h B_{\m'} = \h P^{-1}\g^\n_{\ \m'}\na_\n\h P&(3.18)\cr}
$$
where
$$
\g^\n_{\ \m'}=\{\s^{\m'}_{\ \n}\}^{-1}
\eqno(3.19)
$$
These functions in the case of covariantly constant curvatures (3.12)
can be expressed explicitly in terms of curvatures in a fixed point $x'$.

Introducing a matrix $K=\{K^{\m'}_{\ \n'}\}$ that is a scalar at the point
$x$,
$$
K^{\m'}_{\ \n'}(x,x') = R^{\m'}_{\ \a'\n'\b'}(x')\s^{\a'}(x,x')\s^{\b'}(x,x')
\eqno(3.20)
$$
where $\s^{\m'}=\na^{\m'}\s$ is a tangent vector to the geodesic connecting
the points $x$ and $x'$ at the point $x'$, one can write down the quantities
(3.17) - (3.19) as follows
 [5,6]
$$
\eqalignno{
&\g^\n_{\ \m'} = -g^\n_{\ \a'}
\left({\sin\sqrt K\over \sqrt K}\right)^{\a'}_{\ \m'}&(3.21)\cr
&X^{\m'\n'} =
\left({K \over \sin^2\sqrt K}\right)^{\m'}_{\ \a'}g^{\a'\n'}&(3.21a)\cr
&\h B_{\m'} = -\left({1-\cos\sqrt K\over K}\right)^{\n'}_{\ \m'}\h R_{\n'\a'}
\s^{\a'}&(3.22)\cr}
$$
where $g^\n_{\ \a'}(x,x')$ is a bivector which effects parallel
displacement of vector fields along the geodesic from the point $x'$ to the
point $x$.

The Van Vleck - Morette determinant equals in this case
[5,6]
$$
\Delta=\det\left({\sqrt K\over \sin\sqrt K}\right)
\eqno(3.23)
$$

Mention that if one chooses the normal Riemann coordinates and Fock -
 Schwinger gauge at the point $x'$ for the connection
$$
\eqalignno{
&g^N_{\m\n}(x';x')=\d_{\m\n} \qquad,\qquad
(x^\m-x'^\m)(g^N_{\m\n}(x;x')-\d_{\m\n})=0&(3.23)\cr
&\h A^{FS}_\m(x';x')=0 \qquad,\qquad
(x^\m-x'^\m)\h A^{FS}_\m(x;x')=0&(3.24)\cr}
$$
then
$$
\eqalignno{
&\s^{\m'} = -(x^\m-x'^\m)&(3.25)\cr
&g^\m_{\ \n'} = \d^\m_\n \qquad, \qquad \h P = 1 &(3.26)\cr
&K^{\m}_{\ \n}(x,x') = R^{\m}_{\ \a\n\b}(x')(x^\a-x'^\a)(x^\b-x'^\b)
&(3.27)\cr}
$$
 and one can show that the metric and the connection can be expressed in
 terms of introduced quantities
$$
\eqalignno{
&g^N_{\m\n} = \g^\a_{\ \m'}\g_{\a\n'} =
\left({\sin^2\sqrt K\over K}\right)_{\m\n}&(3.28)\cr
&g_N^{\m\n} = X^{\m'\n'} =
\left({K \over\sin^2\sqrt K}\right)^{\m\n}&(3.28a)\cr
&g_N^{1/2}= \Delta^{-1} = \det\left({\sin\sqrt K\over \sqrt
K}\right)&(3.29)\cr
 &\h A^{FS}_\m = - \h B_{\m'} = -\left({1-\cos\sqrt K\over
 K}\right)^{\n}_{\ \m}\h R_{\n\a} (x^\a-x'^\a)&(3.30)\cr}
 $$

We stress once again that the covariantly constant curvatures
must satisfy also the conditions of integrability (3.13) and (3.14).

Let us make one more important observation. On the covariantly constant
background (3.12), i.e. in symmetric spaces, one can easily solve the Killing
equations
$$
\h L_\xi g_{\m\n} = 2 \na_{(\m}\xi_{\n)} = 0
\eqno(3.31)
$$
Indeed, after differentiating the Killing equations (3.31) and commuting
derivatives it is obtained easily
$$
\na_{(\m}\na_{\n)}\xi^\l=-R^{\l}_{\ (\m|\a|\n)}\xi^\a
\eqno(3.32)
$$
By differentiating this relation and symmetrizing the derivatives we get
$$
\eqalignno{
\na_{(\m_1}\cdots\na_{\m_{2n})}\xi^\l &= (-1)^nR^\l_{\ (\m_1|\a_1|\m_2}
R^{\a_1}_{\ \ \m_3|\a_2|\m_4}\cdots R^{\a_{n-1}}_{\ \
\m_{2n-1}|\a_{n}|\m_{2n})}\xi^{\a_{n}} &(3.33)\cr
\na_{(\m_1}\cdots\na_{\m_{2n+1})}\xi^\l &= (-1)^nR^\l_{\ (\m_1|\a_1|\m_2}
R^{\a_1}_{\ \ \m_3|\a_2|\m_4}\cdots R^{\a_{n-1}}_{\ \
\m_{2n-1}|\a_{n}|\m_{2n}}\na_{\m_{2n+1})}\xi^{\a_{n}} &(3.34)\cr}
$$

Thereby we have found the coefficients of the covariant Taylor series
[21,5,6]
$$
\xi^\m(x) = g^\m_{\ \l'}(x,x')
\sum_{n\ge 0}{(-1)^n\over n!}\sigma^{\mu'_1}\cdots
\sigma^{\mu'_n}\left[\nabla_{(\mu_1}\cdots\nabla_{\mu_n)}\xi^\l\right]_{x=x'}
\eqno(3.44)
$$
 This expansion can be summed up and the Killing vectors of symmetric spaces
 can be written in a closed form
 $$
\xi^\m (x) = g^\m_{\ \n'}(x,x')\left\{(\cos\sqrt K)^{\n'}_{\ \ \l'}
\xi^{\l'}(x') - \left({\sin\sqrt K\over\sqrt K}\right)^{\n'}_{\ \ \l'}
\s^{\rho'}(x,x')(\na_{\rho'}\xi^{\l'})(x')\right\}
\eqno(3.45)
$$
Therefore, all Killing vectors at any point $x$ are determined in terms  of
initial values of the vectors themselves  $\xi^{\l'}(x')$ and their
first derivatives $\na_{\rho'}\xi^{\l'}(x')$ at a fixed point $x'$.  The
maximal number of these parameters is equal to $ d + d(d-1)/2 = d(d+1)/2 $,
since the quantities $\na_{\r'}\xi_{\l'}$ are antisymmetric. The spaces with
maximal number of independent Killing vectors equal to $d(d+1)/2$ are the
spaces of constant curvature and only those.

However, while the initial values of Killing vectors are independent, the
derivatives of those are not. This can be seen immediately if one mentions
that in symmetric spaces they should satisfy also the equation
$$
\h L_\xi R_{\a\b\g\d} = 2 \{ R_{\a\b\s[\d} \na_{\g]}\xi^\s +
	R_{\g\d\s[\b}\na_{\a]}\xi^\s \} = 0
						\eqno(3.46)
$$
This equation plays the role of the integrability conditions for  Killing
equations (3.31) and imposes strict constraints on the possible initial
values of the derivatives of the Killing vectors at the point $x'$.

All the Killing vectors can be splitted in two essentially different sets
$\{P_a\}$ and $\{M_i\}$ according to the values of their initial parameters
$$
\eqalignno{
&P_{\ a}^{\m}(x) = g^\m_{\ \n'}\left(\cos\sqrt K\right)^{\n'}_{\ \ \l'}
P^{\l'}_{\ \ a}(x'), \qquad a = 1,\dots,d; 	&(3.47)\cr
 &M^\m_{\ i}(x) = - g^\m_{\ \n'}
 	\left({\sin\sqrt K\over \sqrt K}\right)^{\n'}_{\ \ \l'} \s^{\r'}
  (\na_{\r'}M^{\l'}_{\ \ i})(x'), \qquad i=1,\dots,p; p\leq {d(d-1)\over 2}
						&(3.48)\cr}
$$
with following initial conditions
$$
\na_{\r'}P^{\l'}_{\ \ a}(x') = 0, \qquad M^{\m'}_{\ \ i}(x') = 0 \eqno(3.49)
$$
and the consistency ones
$$
\eqalignno{
&\det P^{\l'}_{\ \ a}\Big\vert_{x=x'} \ne 0	&(3.50)\cr
&\left\{R_{\a'\b'\s'[\d'} \na_{\g']}M^{\s'}_{\ \ i} + R_{\g'\d'\s'[\b'}
\na_{\a']}M^{\s'}_{\ \ i}\right\}\Bigg\vert_{x=x'}= 0
						&(3.51)\cr}
$$

Using the Killing vectors of symmetric spaces (3.47), (3.48) we define
first-order differential operators of the form
$$
\eqalignno{
&P_{a} = P^\m_{\ a}\na_\m = P^{\n'}_{\ \ a}
\left(\cos\sqrt K\right)^{\m'}_{\ \ \n'}\h D_{\m'} &(3.52)\cr
 &M_{i} = M^\m_{\ i}\na_\m = (\na_{\r'}M^{\m'}_{\ \ i})\s^{\r'}\h D_{\m'}
 						&(3.53)\cr }
$$
where
$$
\h D_{\m'} = \g^\n_{\ \m'}\na_\n
						\eqno(3.54)
$$

One can show that for all the constraints to be fullfieled in symmetric
spaces, including (3.13) and (3.50), the operators (3.52), (3.53) on the
scalar (at the point $x$) fields should generate a Lie algebra
(the algebra of isometries)
$$
\eqalignno{
&[P_a, P_b] = E^i_{\ a b} M_i 			&(3.55)\cr
&[P_a, M_i] = D^b_{\  a i} P_b   		&(3.56)\cr
&[M_i, M_k] = C^l_{\ i k} M_l 			&(3.57)\cr }
$$
with  structure constants satisfying the Jacobi identities
$$
\eqalignno{
&D^a_{\ j [b} E^j_{\ c d]} = 0			&(3.58)\cr
&D^a_{\ i c} D^c_{\ k b} - D^a_{\ k c} D^c_{\ i b} = C^j_{\ i k} D^a_{\ j b}
						&(3.59)\cr
&E^i_{\ a c} D^c_{\ b k} -  E^i_{\ b c} D^c_{\ a k}= E^j_{\ a b} C^i_{\ j k}
						&(3.60)\cr
&C^i_{\ j [k} C^j_{\ m n]} = 0			&(3.61)\cr}
$$
Mention that the generators $M_i$ also form a closed Lie algebra called the
isotropy subalgebra $H$.

In this case using (3.50) and (3.51) one can obtain easily the initial values
of the nonvanishing derivatives of Killing vectors $M$ from (3.56)
$$
\na_{\m'} M^{\n'}_{\ \ i} = D^b_{\ a i} P^{-1 a}_{\ \ \ \ \m'}P^{\n'}_{\ \ b}
						\eqno(3.62)
$$

Then commuting the eq. (3.55) with $P_c$ and using (3.56) we obtain
$$
[P_c, [P_a, P_b]] = D^d_{\ c\, i} E^i_{\ a b} P_d
						\eqno(3.63)
$$
wherefrom one obtains the curvature at the point $x'$ in terms of the
structure constants
$$
R^{\m'}_{\ \ \n'\a'\b'} = D^d_{\ c\, i}E^i_{\ a b} P^{\m'}_{\ \ d}
P^{-1 c}_{\ \ \ \ \n'} P^{-1 a}_{\ \ \ \ \a'}P^{-1 b}_{\ \ \ \ \b'}
						\eqno(3.64)
$$
Taking into account (3.62) and (3.64) one can prove now that, as a
consequence of the Jacobi identities (3.58) - (3.61), the conditions (3.13)
and (3.50) are indeed valid.  This means that giving the structure constants
of a Lie group is equivalent to Riemann curvature tensor.
The structure of the isometry group is completely determined by the
curvature tensor at a fixed point $x'$.
To be more precise the symmetric space is isomorphic to the quotient space
$G/H$ of the isometry group $G$ by the isotropy subgroup $H$
[61].

In more general case when the operators (3.52), (3.53) act on a  multiplet of
fields $\varphi = \{\varphi^A\}$ additional curvature terms arise in
commutation relations (3.55) - (3.57)
$$
\eqalignno{
&[P_a, P_b] = E^i_{\ a b} M_i + \h F_{a b}		&(3.65)\cr
&[P_a, M_i] = D^b_{\  a i} P_b + \h G_{a i}  		&(3.66)\cr
&[M_i, M_k] = C^j_{\ i k} M_j + \h H_{i k}		&(3.67)\cr}
$$
where the following notations for the projections of the curvature of
background connection (2.5) on the Killing vectors are introduced
$$
\eqalignno{
&\h F_{a b} = P^\m_{\ a}P^\n_{\ b}\h R_{\m\n}		&(3.68)\cr
&\h G_{a i} = P^\m_{\ a}M^\n_{\ i}\h R_{\m\n}		&(3.69)\cr
&\h H_{i k} = M^\m_{\ i}M^\n_{\ k}\h R_{\m\n}		&(3.70)\cr}
$$
Generally speaking, one should add also the commutation relations between
operators $P_a, M_i$ and the curvatures $\ F_{ab}, \h G_{ai}, \h H_{ik}$
and between the curvatures themselves  which are the direct consequence
of the conditions (3.12) - (3.14).

Hence we conclude that the formulation of the low-energy approximation in
explicitly covariant way naturally leads to the condition of covariant
constancy of curvatures (3.12) and thereby to the existence of some nontrivial
invariance group with generators  (3.52) and (3.53), the algebra of this
group given by the commutation relations (3.55)-(3.57).

This structure seems to be very useful for calculating the effective potential
of quantum gravity. A simple example how one can make use of this group is
presented in subsect. 4.1. We are going to investigate this point in full
measure in the future.

\bigskip
\bigskip
\centerline{\bf	 IY. APPROACHES FOR CALCULATING THE HEAT KERNEL}
\nobreak
\bigskip
\nobreak
The problem of calculating the low-energy heat kernel and the low-energy
effective action is elaborated not so good. In contrast
to good developed Schwinger - De Witt technique
[9-13,21,5,6],
and also to nonlocal covariant perturbation theory
[48,49],
here only partial success is achieved and various approaches to the
problem are only outlined (see our recent papers [67, 68]).
The general solution of the problem of  explicitly
covariant calculation of the effective action is still not found. That is why
we adduce below a number of approaches that are worth  notice and can
lead in future to a considerable progress.

Thus the problem is  the following. One has to obtain a local covariant
expression that would describe adequately the low-energy limit of the trace
of the heat kernel and that would, being expanded in curvatures, reproduce
all terms without covariant derivatives in the asymptotic expansion of heat
kernel, i.e. the HMDS-coefficients. In other words, this expression can
be called the generating function for HMDS-coefficients. If one finds such an
expression, then one can simply determine the $\zeta$-function (2.16)
and, therefore, the low-energy limit of the effective action (2.14).

\bigskip
\centerline{\sc	 4.1.  Schwinger's method}
\nobreak
\medskip
\nobreak
This is one of the oldest approaches. Nevertheless it turned out to be very
successful in the case of an Abelian gauge field and allowed Schwinger to
obtain in a very elegant way the effective lagrangian of quantum
electrodynamics
[8].
The formulation of the method in case of quantum gravity is due to De Witt
[9-12].

Take an abstract formal Hilbert space with  the basis vectors $|x'>$ that are
eigenvectors of commutative set of Hermitian coordinate operators
$$
 [x^\m,x^\n] = 0 \quad , \quad x^\m|x'> = x'^\m|x'>
\eqno(4.1)
$$
and are normalized according to
$$
<x''|x'> = \d (x'',x')
\eqno(4.2)
$$
The vectors $|x'>$ transform as scalar densities of weight $1/2$
under diffeomorphisms and as fields $\varphi=\{\varphi^A\}$ under gauge
transformations.

Then one introduces Hermitian noncommuting covariant momentum operators
$$
[\Pi_\m,\Pi_\n] = - \h R_{\m\n}
\eqno(4.3)
$$
where $ \h R_{\m\n}(x)$ is an anti-Hermitian function of the  coordinate
operators. The commutation relations of these operators have the form
$$
[x^\m,\Pi_\n] = i \d^\m_\n
\eqno(4.4)
$$
and the matrix elements of the momentum operators are
$$
<x''|\Pi_\m|x'> = -i\na_\m'\d(x'',x')
\eqno(4.5)
$$
The self-adjoint operator (2.8)
$$
H(x,\Pi) = g^{-1/4}F g^{-1/4} =
  g^{-1/4}\Pi_\m g^{1/2}g^{\m\n}\Pi_\n g^{-1/4} + Q + m^2
\eqno(4.6)
$$
is treated naturally as the Hamiltonian of an abstract dynamical
system.

Make an analytic continuation in the complex plane of the variable $t$ on the
positive semi-axis of the imaginary axis, i.e substitute  $t=is$, where
$s>0$, and consider instead of the heat kernel $\exp(-t F)$ the unitary
evolution operator $\exp(-i s H(x,\Pi))$.  The matrix elements of the evolution
operator determine the evolution kernel (or Schr\"odinger kernel)
$$
U(s|x'',x')=<x''|\exp(-isH(x,\Pi))|x'>
\eqno(4.7)
$$
that satisfies the Schr\"odinger equation
$$
i{\partial \over \partial s} U(s|x'',x') = <x''(s)| H(x(s),\Pi(s))|x'>
\eqno(4.8)
$$
where
$$
x(s) = \exp(isH)x\exp(-isH) \quad , \quad \Pi(s) = \exp(isH)\Pi\exp(-isH)
\eqno(4.9)
$$
and
$$
<x''(s)| = <x|\exp(-isH)
\eqno(4.10a)
$$
is the eigenvector of the operator $x(s)$
$$
<x''(s)|x(s) = x''<x''(s)|
\eqno(4.10)
$$
It is obvious, that $H(x(s),\Pi(s)) = H(x,\Pi)$.
The corresponding operator dynamical equations (Heisenberg ones) have the
form
$$
\eqalignno{
{d\,x^\m(s) \over d\,s} &= i[H, x^\m(s)] &(4.11)\cr
{d\,\Pi_\m(s) \over d\,s} &= i[H, \Pi_\m(s)] &(4.12)\cr
}
$$
Using the commutation relations (4.3) and (4.4) it is not difficult to
calculate the commutators in the right-hand side of these equations.
If one manages to integrate these operator equations with initial
conditions  $x(0)=x, \Pi(0)=\Pi $ in the form
$$
\eqalignno{
x(s) &= \exp(is\,Ad_H)x = f_1(s|x,\Pi) &(4.13)\cr
\Pi(s) &= \exp(is\,Ad_H)\Pi = f_2(s|x,\Pi)&(4.14) \cr}
$$
where
$$
Ad_H f = [H,f]
\eqno(4.15)
$$
then one can express the initial momentums in terms of coordinate operators
$$
\Pi = f_3(s|x(s),x)
\eqno(4.16)
$$
put them into the expression for the Hamiltonian (4.6) and order them
using the commutation relations (4.3), (4.4) so that all
$x(s)$ would be placed to the left and all  $x$ to the right.
(It is evident that the coordinate operators at different time $s$ do not
commute.)

As a result the Hamilton operator assumes an air
$$
H(x,\Pi) = {\rm T}\,h(s|x(s),x)
\eqno(4.17)
$$
where the symbol $T$ means the ordering from the right to the left
with respect to  the variable $s$. Finally the Schr\"odinger equation (4.8)
takes the form
$$
i{\partial \over \partial s} U(s|x'',x') = h(s|x'',x')U(s|x'',x')
\eqno(4.18)
$$
that can be integrated easily
$$
U(s|x'',x') = C(x'',x')\exp\left(-is\int\limits^s\,h(s|x'',x')\right)
\eqno(4.19)
$$
The function $C(x'',x')$ should be determined then from the initial condition
$$
U(0|x'',x') = <x''|x'> = \d(x'',x')
\eqno(4.20)
$$
and transformation properties of Schr\"odinger kernel, e.g. from the
relations of the form (4.5). At last one obtains the heat kernel from
Schr\"odinger kernel by doing the inverse analytic continuation $s=-it$.

Consider now the low-energy approximation (3.7). It is sufficient to confine
oneself to covariantly constant curvatures of the background fields (3.12).
The most convenient way to apply the Schwinger's method is to use the Fock -
Schwinger gauge and normal Riemann coordinates at the point $x'$  (3.23),
 (3.24). Then one can express the metric and connection directly in terms  of
 curvatures at the point $x'$ and use the formulae  (3.27)-(3.30).

The covariant momentum operator  $\Pi$ takes the form
$$
\Pi_\m = p_\m - i\h A_\m
\eqno(4.21)
$$
where $p_\m$ are the ordinary Hermitian commuting momentum operators
$$
[p_\m,p_\n] = 0 \quad , \quad [x^\m,p_\n] = i\d^\m_\n
\eqno(4.22)
$$
All commutators at coinciding points $x''=x'$ (in particular, the function
$h(s|x',x')$ and, therefore, the Schr\"odinger kernel at coinciding points
$U(s|x',x')=<x'(s)|x'>$ ) are expressed in covariant way only in
terms of curvatures.

However, one manages to solve the operator Heisenberg equations (4.11),
(4.12) only in exceptional cases. Let us confine ourselves to the covariantly
constant gauge field in flat space.

In this case we have
$$
\eqalignno{
&R_{\m\n\a\b}=0&\cr
&[\Pi_\a,g_{\m\n}] = [\Pi_\a,Q] = [\Pi_\a,\h R_{\m\n}] =
[\h R_{\m\n},\h R_{\a\b}] = [Q,\h R_{\m\n}] = 0&(4.23)\cr}
$$
and Heisenberg equations take the form
$$
\eqalignno{
 {dx^\m\over ds} &= -2g^{\m\n}\Pi_\n &(4.24)\cr
{d\Pi_\m\over ds} &=-2i\h R_{\m\a}g^{\a\n} \Pi_\n&(4.25)\cr}
$$
These equations can be easily solved
$$
\Pi(s) = g\exp(2isg^{-1}\h R)g^{-1}\Pi \quad,
\quad x(s) = x + {\exp(2isg^{-1}\h R)-1 \over ig^{-1}\h R}g^{-1}\Pi
\eqno(4.26)
$$
where $g=\{g_{\m\n}\}, \h R=\{\h R_{\m\n}\}$.
Hence
$$
\Pi = {i\h R \over \exp(2isg^{-1}\h R)-1}(x(s)-x)
\eqno(4.27)
$$
and then
$$
H = \Pi g^{-1}\Pi+ Q + m^2 = {1\over 4}(x(s)-x)\left({\h Rg^{-1}\h R \over
\sin^2(sg^{-1}\h R)}\right)(x(s)-x) + Q + m^2
\eqno(4.28)
$$
Finally, using the commutator
$$
[x(s),x] =
- {\exp(2isg^{-1}\h R)-1 \over g^{-1}\h R}g^{-1}
\eqno(4.29)
$$
for ordering the operators $x(s)$ and $x$ we get
$$
\eqalignno{
h(s|x'',x') &= <x''(s)|H|x'> &\cr
&= {1\over 4}(x''-x')
\left({\h R g^{-1}\h R \over \sin^2(sg^{-1}\h R)}\right)(x''-x')
-{i\over 2}{\rm Sp} (g^{-1}\h R \cot (sg^{-1}\h R)) + Q + m^2&\cr
& &(4.30)\cr}
$$
where $Sp$ denotes the trace over vector indices. The solution of the
Schr\"odinger equation with initial condition (4.20) take now the form
$$
\eqalignno{
U(t|x'',x') &= (4\pi t)^{-d/2}g^{1/2}\h P(x'',x')
\det\left({tg^{-1}\h R \over \sinh(tg^{-1}\h R)}\right)^{1/2}&\cr
&\times \exp\left(-t(m^2 + Q) -
{1\over 4}(x''-x')^\m(\h R\coth(tg^{-1}\h R))_{\m\n}
(x''-x')^\n \right)&(4.31)\cr}
$$
where we came back to the variable $t=is$ for convenience of further
references, $\det$ means the determinant over vector indices and $\h P(x,x')$
is the parallel displacement operator.  This factor is necessary to
ensure the proper covariant transformation properties of the Schr\"odinger
kernel under gauge transformations.

 This solution for the Abelian gauge group $U(1)$ (quantum electrodynamics)
 was obtained first by Schwinger
[8].
However, as we see, it is valid in the much more general case of arbitrary
semisimple gauge group and covariantly constant background fields in flat
space
$$
\eqalignno{
&R_{\m\n\a\b} = 0&\cr
&\na_\m \h R_{\a\b} = \na_\m Q = 0&(4.32)\cr}
$$
The consistency conditions (3.14), (3.15) for these equations take in this
case the form
$$
[\h R_{\m\n},\h R_{\a\b}] = [\h R_{\m\n},Q] = 0
\eqno(4.33)
$$
This means that the curvature and the potential term take their values only
in the Cartan subalgebra of the algebra of the gauge group, the number of
independent components being equal to dimension of the Cartan subalgebra,
i.e. the rank of the algebra. Hence this is a nontrivial extension of the
Schwinger result to the non-Abelian gauge group.

We shall apply this result below to the calculation of the effective
potential in a Yang-Mills model with matter fields.

Notwithstanding all the elegance the Schwinger's method is not  widely
adapted for the present in general case of nontrivial curved background
because of the extreme complexity of the nonlinear operator equations (4.11)
and (4.12).

\bigskip
\centerline{\sc	 4.2.  Fourier integral method}
\medskip

This method is based on the formal representation of the heat kernel in the
form
(2.7)
$$
U(t|x,x') = \exp(-tH)\h P(x,x')\d(x,x')
\eqno(4.34)
$$
Using the representation of the $\d$-function in form of a covariant Fourier
integral
[5,6]
$$
\d(x,x') = g^{1/4}(x)g^{-1/4}(x')\Delta^{1/2}(x,x')
\int{dk_{\m'}\over (2\pi)^d}\exp(ik_{\m'}\s^{\m'}(x,x'))
\eqno(4.35)
$$
we get
$$
\eqalignno{
U(t|x,x') =& g^{1/4}(x)g^{-1/4}(x')\Delta^{1/2}(x,x')\h P(x,x')&\cr
&\times\exp(-tm^2)
\int{dk_{\m'}\over (2\pi)^d} \exp(ik_{\m'}\s^{\m'}(x,x'))\Phi(t|k,x,x')
&(4.36)\cr}
$$
where
$$
\eqalignno{
&\Phi(t|k,x,x') = \exp(-tA)\cdot 1&(4.37)\cr
&A=-\exp(-ik_{\m'}\s^{\m'})\h P^{-1}\Delta^{-1/2}\sq\Delta^{1/2}\h P
\exp(ik_{\m'}\s^{\m'}) + \h P^{-1}Q\h P&(4.38)\cr}
$$

Introduce now
[5,6]
new operators of covariant derivatives as follows
$$
\h D_{\m'}=\g^\n_{\ \m'}\na_\n
\eqno(4.39)
$$
These operators appear to be very convenient since:  i) they commute being
applied to scalars at the point $x$ and ii) the commutators between them and
the tangent vectors $\s^{\m'}$ have the usual very simple form
$$
\left[\h D_{\m'}, \s^{\n'}\right] = \d^{\n'}_{\ \m'}
\eqno(4.40)
$$
Subsequently the operators $\h D_{\m'}$ and the vectors  $\s^{\n'}$ play the
role of usual derivatives and coordinates. It is this fact that makes it
possible to construct the covariant Fourier integral in the form (4.35).

Using the representation of the Laplace operator in terms of $\h D_{\m'}$,
[5,6]
$$
\sq = \Delta\h D_{\m'}\Delta^{-1}X^{\m'\n'}\h D_{\n'}
\eqno(4.41)
$$
where $X^{\m'\n'}$ is given by (3.17a) and recalling the definition of the
quantity $\h B_{\m'}$ (3.18) yields
$$
\eqalignno{
A &= -\Delta^{1/2}\left(\h D_{\m'}
+\h B_{\m'}+ik_{\m'}\right)\Delta^{-1}X^{\m'\n'}\left(\h D_{\n'}
+\h B_{\n'}+ik_{\n'}\right)\Delta^{1/2} + \h P^{-1}Q\h P &\cr
&= -\Delta^{1/2}\left(\h D_{\m'}
+\h B_{\m'}\right)\Delta^{-1}X^{\m'\n'}\left(\h D_{\n'}
+\h B_{\n'}\right)\Delta^{1/2} + \h P^{-1}Q\h P &\cr
&- ik_{\m'}\left(X^{\m'\n'}\h D_{\n'}+\h D_{\n'}X^{\m'\n'}
+2X^{\m'\n'}\h B_{\n'}\right) + k_{\m'}k_{\n'}X^{\m'\n'}&(4.42)\cr}
$$

Thus the function  $\Phi$  is presented as the result of the  action of the
exponent of the differential operator $A$ on the identity. Moving the
derivatives to the right until they act on the identity and give zero the
final result will be expressed in terms of derivatives of the two-point
functions involved ($X^{\m'\n'}, \ \Delta, \h B_{\m'}$ and $\h P^{-1}Q\h P$).
All these functions are scalars at the point $x$ and therefore the operators
 $\h D_{\m'}$ commute when applied to them.

These expressions give the heat kernel in general case of arbitrary
background fields. In low-energy approximation one can make use of
 the closed expressions for the functions $X^{\m'\n'}, \ \Delta$ and
$\h B_{\m'}$ in terms of curvatures (3.21)-(3.23) and use also  the fact that
the potential term commutes with derivatives
$$
\left[\h D_{\m'}, \h P^{-1}Q\h P\right] = 0
\eqno(4.43)
$$

As a consequence the coincidence limit (the diagonal) of the heat kernel
$$
U(t|x,x) = \exp(-tm^2)\int{dk\over (2\pi)^d} \Phi(t|k,x,x)
\eqno(4.44)
$$
will depend on $x$ only through the trivial factor $g^{1/2}(x)$ and will be
expressed manifestly covariant only in terms of curvatures.

Changing the integration variable $k\to k/\sqrt t$ one can rewrite the heat
kernel in coinciding points as
$$
U(t|x,x) = (4\pi t)^{-d/2}\exp(-t(m^2+Q))g^{1/2}(x)\Omega(t|x,x)
\eqno(4.45)
$$
where
$$
\eqalignno{
\Omega(t|x,x) = \lim_{x\to x'}\int{dk_{\m'}\over \pi^{d/2}}g^{-1/2}
&\exp\Biggl\{-k_{\m'}k_{\n'}X^{\m'\n'}
+i\sqrt t k_{\m'}\left(X^{\m'\n'}\h D_{\n'}+\h D_{\n'}X^{\m'\n'}
+2X^{\m'\n'}\h B_{\n'}\right) &\cr
&+t\Delta^{1/2}\left(\h D_{\m'}+\h B_{\m'}\right)
\Delta^{-1}X^{\m'\n'}\left(\h D_{\n'}+\h B_{\n'}\right)\Delta^{1/2}
\Biggr\}\cdot 1&(4.46)\cr}
$$
By the way, this formula can be regarded as a generating function for
HMDS-coefficients in case of covariantly constant background fields.
Indeed, introducing an  averaging over momentums at the point $x'$ with a
Gaussian measure  defined by
$$
\eqalignno{
&<\Phi(k)> =  \lim_{x\to x'}\int{dk\over \pi^{d/2}}g^{-1/2}
\exp(-g^{\m'\n'}(x')k_{\m'} k_{\n'})\Phi(k)&(4.47)    \cr
&<1> = 1 \qquad,\qquad <k_\m> = 0 \qquad,\qquad
 <k_\m k_\n> = {1\over 2}g_{\m\n}&\cr
&<k_{\m_1}\cdots k_{\m_{2n+1}}> = 0 ,&\cr
&<k_{\m_1}\cdots k_{\m_{2n}}> = {(2n)!\over 2^{2n}n!}g_{(\m_1\m_2}\cdots
g_{\m_{2n-1}\m_{2n})}&(4.47a)\cr}
$$
and denoting
$$
\eqalignno{
A_0 &= \Delta^{1/2}\left(\h D_{\m'}+\h B_{\m'}\right)
\Delta^{-1}X^{\m'\n'}\left(\h D_{\n'}+\h B_{\n'}\right)\Delta^{1/2}&(4.48a)\cr
A_1 &= i k_{\m'}\left(X^{\m'\n'}\h D_{\n'}+\h D_{\n'}X^{\m'\n'}
+2X^{\m'\n'}\h B_{\n'}\right) &(4.48b)\cr
A_2 &= - k_{\m'}k_{\n'}(X^{\m'\n'}-g^{\m'\n'}(x'))&(4.48c)\cr}
$$
it is  obtained
$$
\Omega(t|x,x) = <\exp(A_2 + \sqrt t A_1 + tA_0)\cdot 1>
\eqno(4.49)
$$
Hence for the coefficients of asymptotic expansion (HMDS-coefficients) (2.11)
$$
\Omega(t|x,x) = \sum_{k \ge 0}{(-t)^k\over k!} a_k(x,x)
\eqno(4.50)
$$
one finds
$$
\eqalignno{
a_{k}(x,x) &= (-1)^k\sum_{N \ge 0}{k!\over N!}
\sum_{\textstyle{0\le k_1,...,k_N\le 2 \atop k_1+\cdots +k_N=k}}
<A_{k_1}\cdots A_{k_N} \cdot 1>&(4.51)\cr}
$$
Here one should move all derivatives to the right (to identity) and all
functions - to the left and put $x=x'$. As a result, we will come to an
expression in terms of derivatives of the functions $X^{\m'\n'}, \Delta$ and
$\h B_{\m'}$ (3.21)-(3.23) at the coinciding points $x=x'$ that are
expressed in turn only through curvatures.

{}From dimensional grounds it is evident that the coefficients of odd
order vanish
$$
a_{2k+1}(x,x) = 0
\eqno(4.52)
$$
and those of even order $a_{2k}(x,x)$ are built from invariants of
$k$-th order in curvature.

\bigskip
\centerline{\sc	 4.3. Algebraic approach}
\medskip

There exist a very elegant possibility to construct the heat kernel using
only the commutation relations of covariant differential operators [67].

The idea consists in the following.  Take a set of operators $P_a$
 forming a closed, actually nilpotent, Lie algebra
$$
\eqalignno{
[P_a, P_b]  &= \h F_{ab}			&(4.53)\cr
[P_c, \h F_{ab}] &= 0				&(4.54)\cr}
$$
This algebra is substantially a subalgebra of the general algebra (3.65) -
(3.67) considered in previous section  in the case of covariantly constant
curvature (3.12)  in the  flat space case $E^i_{\ a b} = 0$.

Consider an element of corresponding Lie group in exponential parametrization
$$
\Psi(k) = \exp(\sqrt t k^a P_a)
\eqno(4.55)
$$
where $t, k^a$ are the parameters and the factor  $\sqrt t$ is introduced
for convenience of further computations.

Define a quantity $Z(t)$ by averaging the group element over Gaussian
measure
$$
\eqalignno{
Z(t) &= < \Psi(k) >&\cr
     &\equiv \int\,dk\g^{1/2}(4\pi)^{-d/2}\exp\left(-{1\over 4}\g_{ab}k^a
        k^b\right) \Psi(k)&(4.56)\cr}
$$
so that $<1>=1$.
Here we introduce a symmetric non-degenerate positive definite matrix
$\g_{ab}, \g=\det\g_{ab}$ that commutes with operators $P_a$
$$
[P_c, \g_{ab}] = 0
\eqno(4.57)
$$

Let us find a differential equation that is satisfied by  $Z(t)$.
We have
$$
{\partial\over\partial t}Z(t) = {1\over 2\sqrt t}P_a<k^a\Psi(k)>
\eqno(4.58)
$$
Then integrating by parts and using the properties of the Gaussian
measure (4.56), it is easy to show
$$
<k^a\Psi(k)> = 2 \g^{ab}<{\partial\over \partial k^b}\Psi(k)>
\eqno(4.59)
$$
where $\g^{ab}$ is the matrix inverse to $\g_{ab}$.

 Now calculate the derivative $\partial/\partial k^a \Psi$. To do this
consider the operator
$$
A_a = \Psi^{-1}{\partial\over\partial k^a}\Psi
\eqno(4.60)
$$
It satisfies the equation
$$
{\partial\over \partial\sqrt t}A_a = - [k^b P_b, A_a]
\eqno(4.61)
$$
and the initial condition
$$
A_a\big\vert_{t=0} = {\partial\over \partial k^a}
\eqno(4.62)
$$
Hence
$$
A_a = \exp\left(-\sqrt t\,Ad_{k\h D}\right){\partial\over\partial k^a}
\eqno(4.63)
$$
where the operator $Ad_{k\h D}$ is defined according to
$$
Ad_{k\h D} B = [k^a P_a, B]
\eqno(4.64)
$$
Computing the commutators it is easy to show finally
$$
\eqalignno{
Ad_{k\h D}{\partial\over \partial k^a} &= - P_a 		&(4.65a)\cr
(Ad_{k\h D})^2{\partial\over \partial k^a} &= -  k^b\h F_{ba} 	&(4.65b)\cr
(Ad_{k\h D})^n{\partial\over \partial k^a} &= 0 \qquad, \qquad (n\ge 3)
								&(4.65c)\cr}
$$
Thus we obtain from (4.63)
$$
A_a = {\partial\over\partial k^a} + \sqrt t P_a + {1\over 2}t\h
F_{ab}k^b
							\eqno(4.66)
$$
and therefore
$$
\eqalignno{
{\partial\over\partial k^a}\Psi &=
\Psi(\sqrt t P_a + {1\over 2}t\h F_{ab}k^b) &\cr
   &= (\sqrt t P_a - {1\over 2}t\h F_{ab}k^b) \Psi	&(4.67)\cr}
$$
where the relation $\Psi(-k)=\Psi^{-1}(k)$ has been used.

By substituting this expression into  the equation (4.59), it is obtained
that
$$
<k^b\Psi> = 2\sqrt t\tilde G^{ba} P_a<\Psi>
						\eqno(4.68)
$$
where  $ \tilde G^{ab}$ is the matrix inverse to
$$
 \tilde G_{ab} = \g_{ab} + t \h F_{ab}
						\eqno(4.69)
$$
If one substitutes now this expression into the equation (4.68),
decomposes the matrix $\tilde G^{ab}$ in its symmetric and
antisymmetric parts and makes use of commutation relations (4.53) then one
finds finally the equation for the quantity $Z(t)$ (4.56)
$$
{\partial\over\partial t}Z(t) = \Big(P_a G^{ab}(t) P_b + M(t)\Big)Z(t)
						\eqno(4.70)
$$
with initial condition
$$
Z(0) = 1
						\eqno(4.71)
$$
where $G^{ab}(t)$ is the matrix inverse to the matrix $G=\{G_{ab}(t)\}$
$$
\eqalignno{
G(t) &=\g - t^2\h F\g^{-1}\h F				&(4.72)\cr
M(t) &= {1\over 2}{\rm Sp}\left({t\g^{-1}\h F\g^{-1}\h F \over 1-
t^2\g^{-1}\h F\g^{-1}\h F}\right)			&(4.73)\cr}
$$
Here and further the matrix notations $\g=\{\g_{ab}\}$ and $\h F=\{\h
F_{ab}\}$  are used and $Sp$ means the usual matrix trace.

 Consider now the commutation relations (4.53) at greater length. Since the
matrix $\h F_{ab}$  is antisymmetric it can be transformed  by orthogonal
matrices to a diagonal block form that have on the diagonal
two-dimensional antisymmetric matrices or zeros. Without loss of generality
one can put
$$
\eqalignno{
\g^{-1}\h F &=
\{\g^{ab}\h F_{bc}\} = {\rm diag}\left( \h F_1\eps, \h F_2\eps, \cdots ,
                       \h F_n\eps, 0,\cdots , 0 \right)
&(4.74)\cr}
$$
where
$
\eps = \left({0 \atop -1 }\ {1 \atop 0 }\right)
$

This means that all operators can be split in two-dimensional sets of
pairwise noncommuting operators $P_i^+$ and $P_i^-$
$$
 [P_i^+,P_k^-] = 0 , \qquad (i\ne k) \qquad, [P_i^+,P_i^-]=\h F_i
						\eqno(4.75)
$$
Together with the center of the algebra (the unit operator) they
form closed three - dimensional subalgebras. In this basis the matrix
$G_{ab}$ (4.72) is diagonal
$$
\eqalignno{
\g^{-1}G(t) =
\{\g^{ab}G_{bc}(t)\} = {\rm diag}\big\{&(1-t^2\h F_1^2), (1-t^2\h
F_1^2), (1-t^2\h F_2^2), (1-t^2\h F_2^2), \cdots , &\cr
  &(1-t^2\h F_n^2), (1-t^2\h F_n^2), 1, \cdots , 1\big\}&(4.76)\cr}
$$
each eigenvalue corresponding to two-dimensional antisymmetric block on the
diagonal of the matrix $\h F_{ab}$ being twofold degenerate.

This leads to the fact that
$$
L(t) = G^{ab}(t)P_a P_b
\eqno(4.77)
$$
at different $t$ commute
$$
[L(t_1), L(t_2)] = 0
\eqno(4.78)
$$
since all noncommuting operators are assembled pairwise in invariant
two-dimensional operators of the form
$$
L_i = (P^+_i)^2 + (P^-_i)^2
\eqno(4.79)
$$
that commute with each other
$$
[L_i, L_k] = 0
\eqno(4.80)
$$

It is obvious that since the commutation relations (4.78) are
invariant they are valid in general case of arbitrary antisymmetric matrix
$\h F_{ab}$ and positive definite matrix $\g_{ab}$.  In other
words the relation
$$
\left[P_a C^{ab}_{(n)}P_b, P_c C^{cd}_{(m)}P_d\right] = 0
\eqno(4.81)
$$
takes place where
$$
C_{(n)} =
\{C^{ab}_{(n)}\} = \left((\g^{-1}\h F)^{2n}\right)\g^{-1}
\eqno(4.82)
$$
Really, the relations (4.81) can be easily proved by making use of the
commutation relations (4.53), the antisymmetry of  $\h F_{ab}$ and
the symmetry of the matrices $C^{ab}_{(n)}$.

Now one can integrate the equation (4.70) and find
$$
Z(t)=\det(1-t^2\g^{-1}\h F\g^{-1}\h F)^{-1/4}\exp\left(tg^{ab}(t)P_a P_b
\right)
						\eqno(4.83)
$$
where $g^{-1}=\{g^{ab}\}$ is the matrix inverse to the matrix
$g=\{g_{ab}\}$
$$
g^{-1}(t)=
\left({{\rm Arth}(t\g^{-1}\h F)\over t\g^{-1}\h F}\right)\g^{-1}
\eqno(4.84)
$$

Thus we have expressed the average of an element of the Lie group over group
parameters with a Gaussian measure (4.56) in terms of the heat kernel for
the Laplace-like operator with the metric $g_{ab}(t)$.

By inverting this procedure, a quite nontrivial result will be obtained.
It is clear that one can choose the matrix $\g_{ab}$ in
arbitrary way, in particular, depending on $t$.  Moreover, this can be done
in such a way that the matrix $g_{ab}$ would not depend on $t$ and would be
a fixed matrix. In other words, let
$$
\g(t) = t\h F \coth(g^{-1}t\h F)
\eqno(4.85)
$$

Then inverting the formula (4.83) with due regard for the definition of the
 quantity $Z(t)$ (4.56) and changing the integration variables  $k \to
 k/\sqrt t$ we prove the

\smallskip
{\sl
{\sc Theorem:}
\smallskip
\nobreak
\smallskip
In the case of nilpotent Lie algebra given by the commutation relations
$$
[P_a, P_b]  = \h F_{ab} \qquad, \qquad
[P_c, \h F_{ab}] = 0
\eqno(4.86)
$$
the heat kernel can be presented in form of an average over the Lie group
with the Gaussian measure
$$
\eqalignno{
\exp(t\sq) =& (4\pi t)^{-d/2}
\det\left({tg^{-1}\h F\over \sinh(tg^{-1}\h F)}\right)^{1/2}&\cr
&\int dk
\exp\left\{-{1\over 4t}k^a(t\h F \coth(g^{-1}t\h F))_{ab}k^b +
k^aP_a\right\}				&(4.87)\cr}
$$
where $\sq = g^{ab} P_a P_b$, $\h F=\{\h F_{ab}\}$ and
$g=\{g_{ab}\}$ is a positive definite symmetric nondegenerate matrix.
 }

\smallskip

Using this representation the heat kernel in flat space in the case of
covariantly constant curvature of the background connection in
coordinate representation can be easily obtained.

We have by definition (2.7)
$$
U(t|x,x') = \exp\left(-t(-\sq+Q+m^2)\right)\h P(x,x')\d(x,x')
\eqno(4.88)
$$

Using the proven theorem we express the heat kernel in terms of the
quantity
$$
\exp(k^aP_a)\h P(x,x')\d(x,x')
\eqno(4.89)
$$
It is easy to show that
[5,6]
$$
\lim_{x\to x'}P_{(a_1} \cdots P_{a_n)} \h P = 0
\eqno(4.90)
$$
Hence
$$
\lim_{x\to x'}(k^aP_a)^n\h P = 0 \qquad,\qquad
\lim_{x\to x'}\left[\exp(k^aP_a), \h P\right] = 0
\eqno(4.91)
$$
and therefore
$$
\exp(k^aP_a)\h P(x,x')\d(x,x')=\h P(x,x')\d(x+k,x')
\eqno(4.92)
$$
(Another way to be convinced of this gauge invariant equation consists in
using the Fock - Schwinger gauge (3.24) when in flat space $\h A_\m=-(1/2)\h
F_{\m\a}(x^\a-x'^\a)$ (3.30).)

Now we substitute (4.87) in (4.88) and use (4.92).
Subsequently the integral over  $k^a$ becomes trivial and one
obtains immediately the heat kernel that coincides (with evident
substitution $\h F_{ab} \to \h R_{\m\n}$), as it should do, with the result
obtained by means of Schwinger's method (4.31).

Thus the main idea consists in the following representation
$$
\exp(t\sq) = \int dk \Phi(t,k|x,x')\exp(k^aX_a)
\eqno(4.93)
$$
where $X_a$ are the generators of a general finite-dimensional Lie group.

If it would be possible to obtain in the general case (3.65) - (3.67) the
formulae similar to (4.87) then it would allow to find the heat kernel in a
number of important cases, in particular, in symmetric spaces, and would lead
finally to the general solution of the effective potential problem in
quantum gravity. This remains still a very interesting problem to solve.

%  end of sect 4

\bigskip
\bigskip
\centerline{\bf	 Y. EFFECTIVE POTENTIAL IN YANG-MILLS THEORY}
\bigskip

Let us now go on to the illustration of the above stated approaches
for the calculation of the effective potential. As a nontrivial
example consider the one-loop effective potential in quantum Yang-Mills model
in flat space with some set of scalar and spinor matter fields (QCD,
GUT models etc.) [68].

The Euclidean action of the model is
$$
S = \int dx\left\{-{1\over 2g^2}{\rm tr}(\h F^2_{\m\n}) +
i\bar\psi(\g^\m\na_\m+M(\varphi))\psi +
{1\over 2}\varphi^T(-\sq)\varphi+V(\varphi)\right\}
						\eqno (5.1)
$$
where
$$
\h F_{\m\n}=\partial_\m\h A_\n - \partial_\n\h A_\m + [\h A_\m,\h A_\n]
\eqno (5.2)
$$
is the strength of gauge (gluon) fields taking the values in the Lie algebra
of the gauge group $G$,  $\h A_\m=A^a_\m C_a$,  $\h
F_{\m\n}=F^a_{\m\n}C_a$, $C_a=\{C^b_{\ ac}\}$ and  $C^b_{\ ac}$ being  the
generators in the adjoint representation and the structure constants of the
gauge group respectively,  $g$ is the interaction coupling constant,
$\varphi=\{\varphi^A\}$ and $\psi=\{\psi^i\}$ are the multiplets of real, for
definiteness, Higgs scalar fields and the Dirac spinor quark ones, which
belong to some, in general, different irreducible representations of the
gauge group, $V(\varphi)$ is a potential for scalar fields and
$M(\varphi)=\{M^i_{\ k}(\varphi)\}$ is a spinor mass matrix,
$\bar\psi=\psi^{\dag} C^T$, the matrix $C$ is defined according to
$\g_\m^T=-C\g_\m C^{-1}$, $\g_\m$ are the Dirac matrices
$\g_{(\m}\g_{\n)}=g_{\m\n}$, $\na_\m=\partial_\m+T(\h A_\m)$ is the covariant
derivative in the representation  $T$ and $\sq=\na_\m^2$.

 Here and everywhere below in this section $tr$ denotes the trace
 only over group indices, other possible indices being left intact.

We limit ourselves to the compact simple gauge group, the generalization
to the semisimple case being trivial, and normalize the generators of the
adjoint representation, i.e. the structure constants, by the condition
$$
{\rm tr}(C_a C_b) =  C^c_{\ ad}C^d_{\ bc} = -2\d_{ab}
\eqno (5.3)
$$

In general setting one should consider the situation when
both the scalar (Higgs) fields and the gauge ones have the background
contributions. (See the discussion in section III.).
The quantization of the model (5.1) in a general covariant gauge leads to
the one-loop effective action
$$
\Gamma_{(1)} = \Gamma_{(1)YM} + \Gamma_{(1)mat}
\eqno (5.4)
$$
Here the contribution of the Yang-Mills fields (and ghosts) proper has the
form
$$
\eqalignno{
&\Gamma_{(1)YM} = {1\over 2}{\rm Tr}\ln\Delta(\l)/\m^2 -
{\rm Tr}\ln F(\l)/\m^2 &(5.5)\cr
&\Delta(\l) = \Delta + \l H &(5.6)\cr
&\Delta^\m_{\ \n} =  -\sq\d^\m_{\ \n} - 2\h F^\m_{\ \n}&(5.7)  \cr
&H^{\m}_{\ \n} = \na^\m\na_\n &(5.8)\cr
}
$$
where $\Delta^\m_{\ \n}(\l)$ is the second order differential operator
determining  the propagator of gauge fields (gluons),
$$
\eqalignno{
& F(\l) = \sqrt{(1-\l)}F & (5.9)\cr
& F= -\sq&(5.10)\cr}
$$
is the operator that determines the ghost propagator, $\l$ is the gauge
fixing parameter and a renormparameter $\m$ is introduced to preserve
dimensions.

 Here and everywhere below we denote by $Tr$ the
 functional trace, viz.  this means that not only the traces over
 all discrete indices (the group ones, the Lorentz ones, etc.) should be
 taken but over the continuous (space-time coordinates) as well. Although we
 use the same notation $Tr$, the explicit meaning of this notation depends on
 the structure of the quantity to which it is applied. In any case the
 functional trace $Tr$ is the trace over {\it all present} indices including
 the continuous ones.

 Although the  factor $\sqrt{(1-\l)}$ in (5.9) seems to be irrelevant, it
 ensures the gauge independence of the {\it regularized} effective  action on
 the mass shell (see the proof below).

The contribution of the matter fields has the form
$$
\eqalignno{
&\Gamma_{(1)mat} = -{\rm Tr}\ln(\g^\m\na_\m+M(\phi))/\m +
{1\over 2}{\rm Tr}\ln N/\m^2 & (5.11)\cr
&N = -\sq+Q(\phi)&(5.11a)\cr}
$$
where $\phi$ is the background Higgs scalar field and the mass matrix of the
Higgs fields $Q=\{Q^A_{\ B}\}$ is of the form
$$
Q^A_{\ B}(\phi) = {\partial^2 \over \partial\phi^B\partial\phi_A}V(\phi)
 							\eqno(5.12)
$$

\bigskip
\centerline{\sc	 5.1. Contribution of gauge fields}
\nobreak
\medskip
\nobreak

The nonvanishing scalar background fields present no special
computing difficulties. But the variety of the forms and possibilities of the
spontaneous symmetry breakdown  obscures the influence of the background
gauge fields themselves. That is why we consider first the most essential
contribution of Yang-Mills fields.

Let us show first of all that the  effective  action (5.5) does not depend on
the gauge parameter $\l$ on the mass shell, i.e.  when the background fields
satisfy the classical equations of motion.

Indeed, by differentiating $\Gamma_{(1)YM}$  with respect to $\l$ and using
the Ward identities
$$
\eqalignno{
\na_\m\Delta^{-1\ \m}_{\ \ \quad \n}(\l)
&= - {1\over 1-\l}\sq^{-1}\big(\na_\n +
J_\m\Delta^{-1\ \m}_{\ \ \quad \n}(\l)\big) &(5.13a) \cr
\Delta^{-1\ \m}_{\ \ \quad \n}(\l)\na^\n &= - {1\over 1-\l}\big(\na^\m +
\Delta^{-1\ \m}_{\ \ \quad \n}(\l)J^\n\big)\sq^{-1}&(5.13b) \cr}
$$
where
$$
J_\m = \na_\n\h F^\n_{\ \m}
\eqno(5.14)
$$
we get
$$
{\partial \Gamma_{(1)YM} \over \partial \l} = {1\over 2(1-\l)^2}{\rm Tr}
\left\{\na_\m J^\m\sq^{-2} + J_\m\Delta^{-1\ \m}_{\ \  \quad \n}
(\l)J^\n\sq^{-2}\right\}
\eqno(5.15)
$$
Hence it is obvious that
$$
{\partial \Gamma_{(1)YM} \over \partial \l}\Bigg\vert_{J=0} = 0
\eqno(5.16)
$$

This leads to the fact that on such background the one-loop effective
action (5.5) does not depend on the gauge fixing condition and coincides,
therefore, with the unique Vilkovisky - De Witt effective action
[13,63,18].

Thus when considering the covariantly constant background fields
satisfying the condition
$$
\na_\m\h F_{\a\b} = 0
\eqno(5.17)
$$
one can simply put $\l=0$  (so called minimal gauge.)

Strictly speaking, our proof needs a substantiation since the
computations above were absolutely formal. As a matter of fact, the
expressions (5.5) and (5.15) for the effective action contain the
ultraviolet divergences that should be regularized at first. One can show,
that the effective action does not depend on the gauge on mass shell ($J=0$),
more consistently using, for example, the $\zeta$ - function  prescription
(2.14) to regularize functional determinants
$$
\Gamma_{(1)YM} = - {1\over 2}\zeta'_{\Delta(\l)}(0) +
                \zeta'_{F(\l)}(0)
\eqno(5.18)
  $$
  Let us calculate the $\zeta$-functions of the operators $\Delta(\l)$ and
  $F(\l)$.  We have from (5.6)
  $$
  \Delta(\l) = \Delta(1) - (1-\l)H
\eqno(5.19)
  $$
    It is easy to show that
  $$
  \Delta^\m_{\ \l}(1)H^\l_{\ \n} = - J^\m\na_\n \quad , \quad
  H^\m_{\ \l}\Delta^\l_{\ \n}(1) = - \na^\m J_\n
\eqno(5.20)
  $$
where  $J_\m$ is given by the formula (5.14).
  That is at $J_\m=0$ the operators $\Delta(1)$ and $H$ are
  orthogonal.  Using these relations we obtain the heat kernel
  $$
  \exp(-t\Delta(\l)) = \exp(-t\Delta(1)) + \exp(t(1-\l)H) - 1
\eqno(5.21)
  $$
  Hence
  $$
  \exp(-t\Delta(0)) = \exp(-t\Delta(1)) + \exp(tH) - 1
\eqno(5.22)
  $$
    which leads finally to
  $$
  \exp(-t\Delta(\l)) = \exp(-t\Delta) + \exp(t(1-\l)H) - \exp(tH)
\eqno(5.23)
  $$
  where $\Delta=\Delta(0)$ is the operator (5.7).

      At last making use of the relation
  $$
  {\rm Tr}(H)^n = {\rm Tr} \sq^n
\eqno(5.24)
  $$
    gives for the trace of the heat kernel
  $$
  {\rm Tr}\exp(-t\Delta(\l)) = {\rm Tr}\exp(-t\Delta) +
  {\rm Tr}\left\{\exp(-t(1-\l)F) - \exp(-t F)\right\}
\eqno(5.25)
  $$
  where $F=-\sq$.

Wherefrom by the definition of the $\zeta$-function (2.16)
  $$
\eqalignno{
  \zeta_{\Delta(\l)}(p) &= \zeta_{\Delta}(p) + ((1-\l)^{-p}-1)\zeta_F(p)
&(5.26)\cr
  \zeta'_{\Delta(\l)}(0) &= \zeta'_{\Delta}(0) - \ln(1-\l)\zeta_F(0)
&(5.27)\cr}
$$
  Then it is evident that
  $$
\eqalignno{
  \zeta_{F(\l)}(p) &= (1-\l)^{-p/2}\zeta_F(p) &(5.28)\cr
  \zeta'_{F(\l)}(0) &= \zeta'_F(0) - {1\over 2}\ln(1-\l)\zeta_F(0)
&(5.29)\cr}
  $$

  As a result the $\l$-depending terms in effective action (5.18) cancel
  exactly and we find
$$
\Gamma_{(1)YM} = - {1\over 2}\zeta'_{\Delta}(0) + \zeta'_{F}(0)
\eqno(5.30)
$$

Thereby we have proven that the regularized effective action (5.18) at
background fields satisfying the condition
$$
J_\m = \na_\n\h F^\n_{\ \m} = 0
\eqno(5.31)
$$
does not depend on the gauge fixing parameter  $\l$.

Let us calculate now the low-energy effective action (effective
potential) in the case of covariantly constant background fields (5.17).
Such field configurations satisfy automatically the condition  (5.31)
$(J_\m=0)$.  Subsequently everything aforesaid about the independence of
the effective action on the gauge is valid.

{}From (5.17) we have immediately
$$
[\h F_{\a\b},\h F_{\m\n}] = 0
\eqno(5.32)
$$
This means that the gauge fields take their values in Cartan subalgebra,
i.e. the nontrivial nonvanishing components of the gauge field exist only in
the direction of the diagonal generators. The maximal number of independent
fields is equal to the dimension of the Cartan subalgebra $r$, i.e. rank of
the group.

It is obvious that operators $\Delta$ (5.7)  and $F$ (5.10) belong
to those considered in previous section and, therefore, all
formulae obtained there are applicable to them.

For the traces of the heat kernels we have from (4.31)
$$
\eqalignno{
{\rm Tr}\exp(-t\Delta)&=\int dx (4\pi t)^{-d/2}{\rm tr}\left\{
\det\left({t\h F\over \sinh(t\h F)}\right)^{1/2}{\rm Sp}\exp(2t\h
F)\right\}&(5.33)\cr
{\rm Tr}\exp(-t F)&=\int dx (4\pi t)^{-d/2}{\rm tr}
\det\left({t\h F\over \sinh(t\h F)}\right)^{1/2}&(5.34)\cr}
 $$
where $\h F=\{\h F_{\m\n}\}$ is here the matrix with both vector and group
indices. The notations  $det$ and $Sp$ mean in present section the
determinant and the trace over vector indices respectively and $tr$ means as
it was stressed above the trace only over the group indices.

Defining the total Yang-Mills $\zeta$-function
$$
\zeta_{YM}(p) = \zeta_{\Delta}(p) - 2\zeta_F(p)
\eqno(5.35)
$$
so that
$$
\Gamma_{(1)YM} = - {1\over 2}\zeta'_{YM}(0)
\eqno(5.35a)
$$
we obtain
$$
\zeta_{YM}(p) =
\int dx (4\pi)^{-d/2}{\m^{2p}\over \Gamma(p)}\int\limits_0^\infty
d t\,t^{p-d/2-1}{\rm tr}\left\{
\det\left({t\h F\over \sinh(t\h F)}\right)^{1/2}\left({\rm Sp}\exp(2t\h F)
- 2\right)\right\}
\eqno(5.36)
$$

 Let us now calculate the traces in formulae (5.33) and(5.34).
Mention, first of all, that the generators of the Cartan subalgebra of
the compact simple group in adjoint representation $C_a, (a=1,\cdots,r)$
(where $r$ is the rank of the group) are the traceless antisymmetric (in
real basis) commuting matrices
$$
[C_a,C_b] = 0 \qquad , \qquad a,b = 1,\cdots,r
\eqno(5.37)
$$
normalized according to the condition (5.3).
Hence they can be diagonalized (in complex basis) simultaneously
$$
C_a = {\rm diag}(0,\cdots,0, i\a_a^{(1)}, -i\a_a^{(1)},
\cdots,i\a_a^{(p)}, -i\a_a^{(p)})
\eqno(5.38)
$$
where {\abf$^{(i)}$} are the positive roots of the algebra, $p=(n-r)/2$ is
the number of positive roots and $n$ is the dimension of the group.  The
number of zeros on the diagonal of the generators of the Cartan subalgebra in
adjoint representation equals the maximum number of commuting generators of
the group, i.e. rank of the group $r$.

Mention also that from the normalization condition (5.3) it follows, in
particular,
$$
\sum_{\a>0}\a_a\a_b
= \d_{ab} \eqno(5.39)
$$
where the sum is to be taken over all positive roots of the algebra.

Therefore for any analytic function of a matrix taking its values in the
Cartan subalgebra in adjoint representation $F=F^aC_a$ we have
$$
{\rm tr}f(F) = r f(0) + \sum_{{\bfm \a}>0}(f(iF{(\a)}) + f(-iF{(\a)}))
\eqno(5.40)
$$
where
$$
F{(\a)} = F^a\a_a
\eqno(5.41)
$$

Using this relation the heat kernels can be rewritten in the form
$$
\eqalignno{
{\rm Tr}\exp(-t\Delta)&=\int dx(4\pi t)^{-d/2}\left\{
rd+2\sum_{\a>0}
\det\left({t F{(\a)}\over \sin(t F{(\a)})}\right)^{1/2}{\rm Sp}\cos(2t
F{(\a)})\right\}&\cr
						&&(5.42)\cr
{\rm Tr}\exp(-t F)&=\int dx (4\pi t)^{-d/2}\left\{r +
2\sum_{{\bfm \a}>0}\det\left({t F{(\a)}\over \sin(t F{(\a)})}\right)^{1/2}
\right\}					&(5.43)\cr}
 $$
where the notation is introduced
$$
F{(\a)} = \{F^a_{\m\n}\a_a\}
\eqno(5.44)
$$
This is still a matrix with vector indices, actually a 2-form.
Now one can calculate the trace and the determinant over the vector
indices. Any 2-form  $F=\{F_{\m\n}\}$  can be  decomposed in a set of
 orthogonal simple 2-forms  $F_{(i)}=\{F_{(i)\m\n}\}$ as follows
$$
\eqalignno{
& F_{\m\n} = \sum_{1\le i\le q}F_{(i)\m\n}H_i&(5.45)\cr
&F_{(i)}F_{(k)} = 0 \qquad, \qquad (i\ne k)&(5.46)\cr
&F_{(i)}^2 = - P_{(i)}  &   \cr}
$$
where $H_i$ are the invariants of the 2-form $F_{\m\n}$, $q \le [d/2]$ is the
 number of nonvanishing independent invariants and $P_{(i)}$ are the
 two-dimensional projectors
 $$
\eqalignno{
P_{(i)}^2 &= P_{(i)}   &       \cr
P_{(i)}P_{(k)} &= 0 \qquad, \qquad (i\ne k)&(5.47)\cr
{\rm Sp}P_{(i)} &= 2                     &\cr}
$$
One can show that in the case  of Euclidean signature all invariants $H_i$
can be regarded to be positive.

Using the expansion (5.45), it is easy to get
$$
\eqalignno{
&{\rm Sp} F^{2k+1} = 0  \qquad, k=1,2,\cdots,      &(5.48)\cr
&{\rm Sp} F^{2k} = (-1)^k 2\sum_{1\le i\le q}H_i^{2k}&(5.49)\cr}
$$
One can consider the $q$  invariants $H_i$ to be determined from
the set of the first $[d/2]$ equations (5.49)  $k=1,2,\cdots,[d/2]$.
Then the rest equations will be satisfied identically.
It is easy to show that $q$ invariants $H_i$ are the positive roots of the
algebraic equation of degree $2[d/2]$
$$
H^{2[d/2]} + c_1H^{2([d/2]-1)} + \cdots + c_{[d/2]-1}H^2 + c_{[d/2]} = 0
\eqno(5.49a)
$$
with the coefficients
$$
c_k = \sum_{1\le j \le k}(-1)^j
\sum_{\textstyle{1\le k_1\le \cdots \le k_j\le k-j+1\atop
 k_1+\cdots+k_j=k}} {1\over k_1\cdots k_j}I_{k_1}\cdots I_{k_j} \eqno(5.49b)
$$
where
$$
I_k = {1\over 2}(-1)^k{\rm Sp} F^{2k}
\eqno(5.49c)
$$

Hence for any analytic function of a 2-form $F=\{F_{\m\n}\}$ it takes place
$$
\eqalignno{
&{\rm Sp}f(F) = (d-2q)f(0) +  \sum_{1\le j\le q}\left(f(iH_j)
+ f(-iH_j)\right)&(5.50)\cr
&\det f(F) = (f(0))^{d-2q}\prod_{1\le j\le q}
f(iH_j)f(-iH_j) \quad,\quad f(0)\ne 0&(5.51)\cr}
$$

By making use of these formulae, one can compute the determinant
over vector indices in (5.42) and (5.43) and obtain finally
$$
\eqalignno{
{\rm Tr}\exp(-t\Delta)=\int dx (4\pi t)^{-d/2}\Biggl\{
&rd + 2\sum_{{\bfm \a}>0}
\prod_{1\le i\le q}\left({tH_i(\a)\over \sinh(tH_i(\a))}\right)&\cr
&\times\left(d + 4\sum_{1\le j\le q}\sinh^2(tH_j(\a))\right)\Biggr\}
&(5.52)\cr
{\rm Tr}\exp(-t F)=\int dx(4\pi t)^{-d/2}\Biggl\{
&r+2\sum_{{\bfm \a}>0}
\prod_{1\le i\le q}\left({tH_i(\a)\over \sinh(tH_i(\a))}\right)\Biggr\}
&(5.53)\cr}
$$
where $H_i(\a)$  are the invariants of the tensor $F_{\m\n}(\a)$ (5.44).

Hence the total $\zeta$-function for gauge fields equals (5.36)
$$
\eqalignno{
\zeta_{YM}(p) =
\int dx (4\pi)^{-d/2}{\m^{2p}\over \Gamma(p)}\int\limits_0^\infty d t\,
&t^{p-d/2-1}\Biggl\{r(d-2) + 2\sum_{{\bfm \a}>0}
\prod_{1\le i\le q}\left({tH_i(\a)\over \sinh(tH_i(\a))}\right)&\cr
&\times\left(d-2 + 4\sum_{1\le j\le q}\sinh^2(tH_j(\a))\right)\Biggr\}
&(5.53a)\cr}
$$
Wherefrom it is immediately seen how the ghost fields effectively decrease
the number of degrees of freedom of gauge field $d\to (d-2)$.

Further calculations are possible only for a concrete gauge group,
i.e. for one or another system of roots of the algebra. For example, for
$SU(N)$ group, as it is well known,
[64],
the rank is $r=N-1$, the dimension is $n=N^2-1$, the number of positive
roots is $p=N(N-1)/2$ and positive roots are given by
$$
\a_{i j} = e_i - e_j \qquad,\qquad (1\le i<j\le N)
\eqno(5.54)
$$
where $e_i$ is an orthonormal basis in $\RR^{N}$.   All roots lie in the
plane orthogonal to the vector  $\sum_{1\le i \le N}e_i$ and are, actually,
$(N-1)$-dimensional vectors in $\RR^{N-1}$.

Let us investigate now the behavior of the heat kernels at $t\to \infty$.
It determines the analytic properties of the corresponding  $\zeta$-function
and one-loop effective action and thereby the stability properties of  the
corresponding vacuum.

{}From (5.52) and (5.53) we have
$$
\eqalignno{
{\rm Tr}\exp(-t\Delta)\sim \int &dx (4\pi)^{-d/2}\Biggl\{rd\,t^{-d/2} +
2^{q+1}t^{q-d/2}\sum_{{\bfm \a}>0}\prod_{1\le i\le q}H_i(\a)&\cr
&\times\sum_{1\le k\le q}\exp\left\{-t\left(\sum_{1\le j\le q}
H_j(\a)-2H_k(\a)\right)\right\}\Biggr\}&(5.55)\cr
{\rm Tr}\exp(-t F)\sim \int &dx (4\pi)^{-d/2}r\,t^{-d/2}&(5.56)\cr}
$$

By the way one can determine from these equations the minimal eigenvalues
of the operators $\Delta$ and $F$. It is easy to show, that
the heat kernel for the Laplace-like second order operator (2.8) on
covariantly constant background (5.17), (5.44), (5.45) behaves at
$t\to\infty$ in general case as follows
$$
{\rm Tr}\exp(-t\Delta) \sim t^{-(d-2q)/2}\exp(-t\l_{min})
\eqno(5.57)
$$
where $(d-2q)$ is the number of dimensions with continuum spectrum when the
eigenvalues are enumerated by both discrete and continuum parameters
$$
\l = \l(p_1,\cdots,p_{d-2q},n_1,\cdots,n_{2q}) \qquad,\qquad
-\infty < p_i < +\infty \qquad,\qquad n_i = 0, 1, 2, \cdots
\eqno(5.58)
$$

The heat kernel for the scalar ghost operator (5.56) behaves good, i.e.
decreases, at $t\to\infty$.  This means that the minimal eigenvalue of the
operator $F$ is equal to zero
$$
\l_{min}(F) = 0
\eqno(5.59)
$$

The heat kernel for vector gluon operator behaves at $t\to\infty$, in
general, not good. This is caused by the self-interaction of the Yang - Mills
fields, viz. by the extremely large value of the gyromagnetic ration of
the gluons (the coefficient in front of $\h F^{\m}_{\ \n}$) in (5.7)) equal
to 2. In this case the second term in (5.55) can be exponentially large.
This is the consequence of the fact, that the operator  $\Delta$ can have, in
general, the negative modes. From (5.55) one can conclude that the minimal
eigenvalue of $\Delta$ is either negative
$$
\l_{min}(\Delta) = - \max_\a\left\{
-\sum_{1\le i\le q}^{\ \quad\prime}H_i(\a) + H_{max}(\a)\right\}
\eqno(5.60)
$$
(where $H_{max}(\a) = \max_{1\le i\le q}H_i(\a)$ is the largest invariant of
 the field $ F_{\m\n}{(\a)}$ (5.44), the prime at the sum meaning that
 the sum does not include this maximal term) or is equal to zero
$$
\l_{min}(\Delta) = 0
\eqno(5.61)
$$
if the previous value (5.60) is positive.

Thus the vector operator $\Delta$ is positive definite and the heat kernel
behaves good at $t\to\infty$ only in the case if the maximal invariant
$H_{max}(\a)$ of the background field $F_{\m\n}{(\a)}$ (5.44) is smaller than
the sum of all other ones.

This is possible only in the case when the number of independent
invariants is equal or greater than two $q\ge 2$, i.e. when the dimension
of the space is not less than four $d\ge 4$.

In case $q=1$, i.e. there is only one independent invariant of the field
$F_{\m\n}{(\a)}$ (this case for the $SU(2)$ group was considered first by
Savvidy
[65]),
without fail the negative modes of the gluon operator exist since the minimal
eigenvalue is negative
$$
\l_{min}(\Delta) = - \max_\a H(\a) < 0
\eqno(5.62)
$$
(in this case  $H(\a)=\sqrt{(F^a_{\m\n}\a_a)^2/2}$).
It is this fact that leads to the well known instability of the Savvidy
vacuum
[66].

Only starting with the number of nonvanishing invariants equal to two ($q=2$)
the instability can disappear. In this case
$$
\l_{min}(\Delta) = - \max_\a\left\{-H_{min}(\a) + H_{max}(\a)\right\}
\eqno(5.63)
$$
where $H_{min}(\a)$ and $H_{max}(\a)$ are the minimal and the maximal
invariants.  Hence it is seen, that if $H_{min}(\a) \ne H_{max}(\a)$, then
again $\l_{min}(\Delta) < 0$. The only possibility to achieve the
absence of the negative modes in the case $q=2$ is to choose these
invariants equal to each other $H_{1}(\a)=H_{2}(\a)=H(\a)$.

Then the heat kernels take the form
$$
\eqalignno{
{\rm Tr}\exp(-t\Delta)&=\int dx (4\pi t)^{-d/2}\left\{
rd + 2\sum_{{\bfm \a}>0}
{t^2H^2(\a)\over \sinh^2(tH(\a))}
\left(d + 8\sinh^2(tH(\a))\right)\right\}&(5.64)\cr
{\rm Tr}\exp(-t F)&=\int dx (4\pi t)^{-d/2}\left\{r +
2\sum_{{\bfm \a}>0}{t^2H^2(\a)\over \sinh^2(tH(\a))}\right\}&(5.65)\cr}
$$
and have both good decreasing as a degree behavior at  $t\to\infty$
$$
\eqalignno{
&{\rm Tr}\exp(-t\Delta)\sim \int dx (4\pi)^{-d/2}16t^{2-d/2}
\sum_{{\bfm \a}>0}H^2(\a)&(5.66)\cr
&{\rm Tr}\exp(-t F)\sim \int dx (4\pi)^{-d/2}r\,t^{-d/2}&(5.67)\cr}
$$
Thus at $q=2, H_1(\a)=H_2(\a)$   the operator  $\Delta$ is positive definite
(except for the zero modes), i.e. there is not instability that is
characteristic to the Savvidy vacuum. It seems on the face of it, that this
case can be realized also in four-dimensional space $d=4$.  However,
as we show below, in case of four dimensions $d=4$ it is not possible to
make the analytic continuation of the equality of two invariants to the
pseudo-Euclidean space of Lorentzian signature, limiting thereby the possible
physical applications of this result.

Consider the case of two invariants in four-dimensional space ($q=2, d=4$) at
greater length.  In this case the invariants $H_i(\a)$ given by the
solutions of the eq.  (5.49) have the simple form
$$
H_{1,2}(\a)=\sqrt{{1\over 2} I_1(\a) \pm {1\over 2}
\sqrt{2 I_2(\a) - I_1^2(\a)}}
						\eqno(5.68)
$$
where $I_k(\a)$ are the  invariants defined by (5.49c), viz.
$$
\eqalignno{
I_1(\a) &= {1\over 2} F_{\m\n}(\a)F_{\m\n}(\a) 	     &(5.68a)\cr
I_2(\a) &= {1\over 2} F_{\m\n}(\a)F_{\n\l}(\a)F_{\l\r}(\a)F_{\r\m}(\a)
							&(5.68b)\cr}
$$
The situation of two equal invariants  $H_1(\a)=H_2(\a)$ in four dimensional
space considered above means, that a relation between the
invariants $I_k$ takes place
$$
I_2(\a) = {1\over 2} I_1^2(\a)
							\eqno(5.69)
$$
But this is possible only in space of Euclidean signature when for any field
$$
I_2(\a) < I_1^2(\a)
							\eqno(5.70)
$$
When going to the pseudo-Euclidean (Lorentzian) signature the sign of
inequality here changes to the opposite one
$$
I_2(\a)  > I_1^2(\a)
							\eqno(5.71)
$$
that leads to the impossibility of the equality condition (5.69) to be
satisfied in Minkowski space.

In Euclidean case both invariants $H_1(\a)$ and $H_2(\a)$ are real, whereas
in Minkowski space one of them is necessarily imaginary.

Only beginning with  $d\ge 5$ there exists such a background that, on  the
one hand, ensures the operator $\Delta$ (5.7) to be  positive definite and,
on the other hand, assumes the analytic continuation on the pseudo-Euclidean
space of Lorentzian signature.

This is a consequence of the general fact that, when doing the analytic
continuation of the results obtained in Euclidean signature to the Lorentzian
signature, one should put, in general,  one of the invariants of any 2-form
to be imaginary, i.e.
$$
H_{q}(\a)=i E(\a)
							\eqno(5.72)
$$

One may call the real invariants  the magnetic and the imaginary the electric
ones. In other words, in Euclidean space all invariants are magnetic, while
in the Minkowski space one of them has to be electric (it can also vanish).

The heat kernel after the substitution $t=is$ becomes the
Schr\"odinger one. The formulae in Minkowski space take the form
$$
\eqalignno{
{\rm Tr}\exp(-is\Delta)=\int &dx (4\pi is)^{-d/2}\Biggl\{rd +
2\sum_{{\bfm \a}>0}\prod_{1\le i\le q-1}
\left({sH_i(\a)\over \sin(sH_i(\a))}\right)
\left({sE(\a)\over \sinh(sE(\a))}\right)         & \cr
&\times\left(d - \sum_{1\le j\le q-1}4\sin^2(sH_j(\a))
+ 4\sinh^2(sE(\a))\right)\Biggr\}			&(5.73)\cr
{\rm Tr}\exp(-i s F)=\int &dx (4\pi is)^{-d/2}\Bigg\{r +
2\sum_{{\bfm \a}>0}
\prod_{1\le i\le q-1}
\left({sH_i(\a)\over \sin(sH_i(\a))}\right)
\left({sE(\a)\over \sinh(sE(\a))}\right)\Bigg\}&\cr
						& & (5.74)\cr}
$$

The presence of the electric field leads to poles in heat kernels,
indeterminacy in integrals over $t$ and, as a consequence, to imaginary part
of the effective potential, i. e. to the creation of the particles and
instability (although not so potent as in the presence of negative modes).

Take notice of the fact, that whereas the heat kernel is good defined,
the Schr\"odinger kernel contains, in general, various poles, divergences
and ambiguities and behaves not good at $s\to \infty$.   That is why
it should be understood as a result of analytic continuation of the
heat kernel, that gives the prescriptions how to get round the  poles in
integrals over $t$ for $\zeta$-function and effective action. When the
electric component is absent such  problems do not appear and the analytic
continuation is trivial.

\bigskip
\bigskip
\centerline{\sc	 5.2. Contribution of matter fields}
\medskip

Consider now the contribution of the matter fields to the one-loop effective
action (5.10). As was stressed above, both the background gauge field
stregth $\h F_{\m\n}$ and the mass matrices $Q(\phi)$ and  $M(\phi)$ are
assumed to be covariantly constant
  $$
 \na_\m\h F_{\a\b} = 0 \qquad,\qquad
 \na_\m M = 0 \qquad,\qquad
 \na_\m Q = 0
 \eqno(5.75)
 $$
that means
$$
[\h F_{\a\b}, \h F_{\m\n}] = 0 \qquad,\qquad
[\h R_{\a\b}, M] = 0 \qquad,\qquad
[\h R_{\a\b}, Q] = 0
 \eqno(5.76)
$$
where $\h R_{\m\n}=T(\h F_{\m\n})=F^a_{\m\n}T_a$ is the curvature and $T_a$
are the generators of the gauge group in the representation to which the
matter fields belong.

For the irreducible representations of compact simple Lie group the
generators $T_a$ are traceless matrices which can be normalized
by the condition [9]
$$
{\rm tr}\,(T_a T_b) = - {D\over n}T^2\d_{ab}
 \eqno(5.77)
$$
where $D$ is the dimension of the representation, $n$ is the dimension of
the group and $T^2=-T_aT_a$ is the eigenvalue of the Casimir operator of
second order for given representation.

The curvature $\h R_{\m\n}$ determines the commutator of the
covariant derivatives on the matter fields:
$$
[\na_\m, \na_\n]\psi = \h R_{\m\n} \psi
 \eqno(5.78)
$$
and analogously for scalar fields.

We do not make here any difference in notations between representations
realized by scalar and spinor fields, as it can not lead to  misunderstanding.

Further, if the mass matrix of fermions $M$ does not contain the Dirac
matrices, as it is assumed, or contains only even number of them, then it
is easy to show, that the determinant of the Dirac operator does not depend
on the sign in front of the derivative and, therefore, it can be expressed in
terms of the squared Dirac operator
$$
\eqalignno{
&{\rm Tr}\ln(\g^\m\na_\m+M)=  {1\over 2}{\rm Tr}\ln\,K&(5.79)\cr
&K = (\g^\m\na_\m+M)(-\g^\n\na_\n+M) =
-\sq - {1\over 2}\g^{\m\n}\h R_{\m\n} + M^2&(5.80)\cr}
 $$
where $\g_{\m\n}=\g_{[\m}\g_{\n]}$ are the generators of the orthogonal group
(Lorentz group) and it has been made use of the relation $[M,\g^\m\na_\m]=0$.

The squared Dirac operator $K$ is now of the Laplace type considered in
previous section. Using the formulae (4.31) one can write down the heat
kernels for the matter fields
$$
\eqalignno{
{\rm Tr}\exp(-t K)&=\int dx (4\pi t)^{-d/2}{\rm tr}\left\{
\exp(-tM^2)\det\left({t\h R\over \sinh(t\h R)}\right)^{1/2}
{\rm tr_\g}\exp\left({1\over
2}t \g^{\m\n}\h R_{\m\n}\right)\right\}&\cr
&&(5.82)\cr
{\rm Tr}\exp(-tN)&=\int dx (4\pi t)^{-d/2}{\rm tr}\left\{\exp(-t Q)
\det\left({t\h R\over \sinh(t\h R)}\right)^{1/2}\right\}&(5.82)\cr}
$$
where $\h R=\{F^a_{\m\n}T_a\}$ is a matrix with vector and group indices,
$tr_\g$ denote the trace over the spinor indices and, as above,
$det$ and $tr$ mean the determinant over vector indices and the trace
over the group ones.

As above the background fields lie in the Cartan subalgebra, i.e.
$$
[T_a,T_b] = 0 \qquad , \qquad a,b = 1,\cdots,r
\eqno(5.83)
$$
and, therefore, the generators $T_a$ can be diagonalized simultaneously
$$
T_a = {\rm diag} (i\n_a^{(1)},\cdots,i\n_a^{(D)})
\qquad , \qquad a = 1,\cdots,r
\eqno(5.84)
$$
where \nbf \ are the weights of the representation $T$, some of them
being, in general,  multiple or equal to zero. They can be expressed in
terms of the highest weight $\m$ and simple roots of the algebra $\l_i$
[64]
$$
\n = \m - \sum_i n_i\l_i
\eqno(5.85)
$$
where  $n_i$ are some non-negative integer coefficients. The highest
weight of the representation characterizes uniquely the representation and
is given by integer numbers
[64]
$$
m_i = 2{(\l_i,\m)\over(\l_i,\l_i)}\qquad,\qquad i=1,\cdots,r
\eqno(5.86)
$$
  For example, the dimension of the representation with the highest weight
  \mbf \ is given by the Weyl formula
$$
D = {\rm dim} T = \prod_{\a>0}{(\a,\chi)\over (\a,\rho)}
\eqno(5.87)
$$
where
$$
\chi=\m+{1\over 2}\sum_{\a>0}\a
\eqno(5.88)
$$
{}From the normalization condition (5.77) we have, in particular,
$$
\eqalignno{
&\sum_{\n\ne 0} d_\n = D - R&(5.89)\cr
&\sum_{\n\ne 0} d_\n\n = 0&(5.90)\cr
&\sum_{\n\ne 0} d_\n \n_a\n_b = {D\over n}T^2\d_{ab}&(5.91)\cr}
$$
Here and further in such formulae the summation is to be taken over all
nonvanishing weights, $d_\n$ are the multiplicities of the weights and  $R$
is the number of zero weights, i.e. the multiplicity of the zero weight.

Now it is easy to obtain for the trace over the group indices of any analytic
function of the matrix $\h R$
$$
{\rm tr} f(\h R) = f(0)R + \sum_{\n\ne 0} d_\n f(i\h F(\n))
\eqno(5.92)
$$
where it is denoted $F{(\n)} = \{F^a_{\m\n}\n_a\}$.

By isolating in mass matrices the singlet contributions
$$
\eqalignno{
&M^2 = M^2_{(0)} + M^{2}_aT_a\qquad,\qquad Q = Q_{(0)} + Q^a  T_a&
(5.93)\cr}
$$
and denoting
$$
\eqalignno{
&M^2{(\n)} = M^{2}_a\n_a \qquad,\qquad Q(\n) = Q^a\n_a&(5.94)\cr}
$$
we get from this for the heat kernels
$$
\eqalignno{
{\rm Tr}\exp(-t K)&=\int dx (4\pi t)^{-d/2}\exp(-tM^2_{(0)})\Biggl\{
R2^{[d/2]} + \sum_{\n\ne 0} d_\n \exp(-i t M^2(\n)) &\cr
&\times\det\left({t F{(\n)}\over \sin(t F{(\n)})}\right)^{1/2}
{\rm tr_\g}\exp\left({i\over 2}t \g^{\m\n} F_{\m\n}{(\n)}\right)\Biggr\}
						&(5.95)\cr
{\rm Tr}\exp(-tN)&=\int dx (4\pi t)^{-d/2}\exp(-tQ_{(0)})\left\{
R + \sum_{\n\ne 0} d_\n
\exp(-i t Q(\n))
\det\left({t F{(\n)}\over \sin(t F{(\n)})}\right)^{1/2}\right\}&\cr
						&&(5.96)\cr}
$$
The determinant over vector indices is to be calculated as before with making
use of (5.51) and is equal to
$$
\det\left({t F{(\n)}\over \sin(t F{(\n)})}\right)^{1/2}=
\prod_{1\le i\le q}\left({tH_i(\n)\over \sinh(tH_i(\n))}\right)
\eqno(5.97)
$$
where $H_i(\n)$ are the invariants of the 2-form
$ F_{\a\b}(\n)= F^a_{\a\b}\n_a$ defined from the equations (5.49).

Finally, we calculate the trace over spinor indices. Using the representation
of a 2-form $F_{\m\n}$  in form (5.45) and the orthogonality of the basic
2-forms $F_{(i)}$ (5.46), one can show, that matrices $\g^{\m\n}F_{(i)\m\n}$
at different $i$ commute
$$
\left[\g^{\m\n}F_{(i)\m\n}, \g^{\a\b}F_{(k)\a\b}\right] = 0
\eqno(5.98)
$$
Hence
$$
\exp\left({i\over 2}t \g^{\m\n}F_{\m\n}\right)
=\prod_{1\le i\le q}\exp\left({i\over 2}t
H_i\g^{\m\n}F_{(i)\m\n}\right)
\eqno(5.99)
$$
The tensors $F_{(i)\m\n}$ are simple 2-forms. This means, that they behave
effectively like the two-dimensional ones.  Therefore, it is valid for
them, in particular, the relation
$$
F_{(i)[\m\n}F_{(i)\a\b]} = 0\qquad\qquad
({\rm no \ summation \ over}\  i!)
\eqno(5.100)
$$
Hence one can obtain using (5.47)
$$
\eqalignno{
&(\g^{\m\n}F_{(i)\m\n})^2 = -4&\cr
&(\g^{\m\n}F_{(i)\m\n})^{2k} = (2i)^{2k}&(5.101)\cr
&(\g^{\m\n}F_{(i)\m\n})^{2k+1} =
(2i)^{2k}\g^{\m\n}F_{(i)\m\n}&\cr}
 $$
i.e. for any analytic function $f(z)$ one has
$$
f\left(\g^{\m\n}F_{(i)\m\n}\right) = {1\over 2}(f(2i) + f(-2i)) + {1\over 4i}
(f(2i) - f(-2i))\g^{\m\n}F_{(i)\m\n}
\eqno(5.102)
$$
This yields
$$
\exp\left\{{i\over 2}t\g^{\m\n}F_{(i)\m\n}H_i\right\} =
\cosh\left(tH_i\right) + {i\over
2}\g^{\m\n}F_{(i)\m\n}\sinh\left(t H_i\right)
\eqno(5.103)
$$

Finally, using again the orthogonality of the 2-forms $F_{(i)\m\n}$ at
different $i$, we find
$$
\eqalignno{
&{\rm tr_\g}\left(\g^{\m\n}F_{(i)\m\n}\g^{\a\b}
F_{(k)\a\b}\right) = 0
\qquad i\ne k&\cr
&{\rm
tr_\g}\left(\g^{\m\n}F_{(i)\m\n}\g^{\a\b}F_{(k)\a\b}
\cdots\g^{\rho\s}F_{(l)\rho\s}\right) \ne 0
\qquad ({\rm only \ if } \ i=k=\cdots=l) &(5.104)\cr}
$$
Therefore, the trace over the spinor indices is completely factorized and we
get
$$
{\rm tr_\g}\exp\left\{{i\over 2}t\g^{\m\n} F_{\m\n}(\n)\right\}
= 2^{[d/2]}\prod_{1\le i\le q}\cosh\left(tH_i(\n)\right)
						\eqno(5.105)
$$

 Taking into account (5.97) and (5.105), the final result for the  heat
 kernel of the spinor field takes the form
 $$
\eqalignno{
{\rm Tr}\exp(-t K)&=\int dx (4\pi t)^{-d/2}\exp(-tM^2_{(0)})2^{[d/2]}
\Biggl\{R + \sum_{\n\ne 0} d_\n
\exp(-i t M^2(\n)) &\cr&\times\prod_{1\le i \le q}\left(tH_i{(\n)}
\coth\left(tH_i{(\n)}\right)\right)\Biggr\}&(5.106)\cr}
$$

The heat kernel for scalar operator (5.96) has the similar form
$$
\eqalignno{
{\rm Tr}\exp(-tN) & =\int dx (4\pi t)^{-d/2}\exp(-tQ_{(0)})\Biggl\{
R + \sum_{\n\ne 0} d_\n\exp(-i t Q(\n))&\cr
&\times\prod_{1\le i\le q}
\left(
{tH_i(\n)\over \sinh
\left(tH_i(\n)\right)}
\right)\Biggr\}
&(5.107)\cr}
 $$

In contrast to the contribution of the gauge fields (5.52) and (5.53))
these heat kernels (5.106) and (5.107) have good exponentially decreasing
behavior at $t\to \infty$ (when some positive singlet contributions in mass
matrices $M^2_{(0)}$ and $Q_{(0)}$ are present). When the singlet
contributions are zero or even negative then the instability appear, that
leads to a reconstruction of the vacuum and to other values of the background
fields ensuring the stability of the vacuum state.

The corresponding formulae for Schr\"odinger kernel in Minkowski space (of
Lorentzian signature), when one of the invariants of the background gauge
field is imaginary (5.72), have the form
 $$
\eqalignno{
{\rm Tr}\exp(-i s K)&=\int dx (4\pi is)^{-d/2}\exp(-isM^2_{(0)})2^{[d/2]}
\Biggl\{R + \sum_{\n\ne 0} d_\n
\exp(sM^2(\n)) &\cr
&\times
sE{(\n)}\coth\left(sE{(\n)}\right)
\prod_{1\le i \le q-1}\left(sH_i{(\n)}
\cot\left(sH_i{(\n)}\right)\right)\Biggr\}&(5.108)\cr
 {\rm Tr}\exp(-i s N)&=\int dx (4\pi is)^{-d/2}\exp(-isQ_{(0)})\Biggl\{
R + \sum_{\n\ne 0} d_\n \exp(sQ(\n)) &\cr&
\times sE{(\n)}\coth\left(sE{(\n)}\right)
\prod_{1\le i \le q-1}\left(sH_i{(\n)}
\cot\left(sH_i{(\n)}\right)\right)\Biggr\}&(5.109)\cr}
 $$

It is appropriate to make here the remarks concerning the analytic
continuation of the heat kernel made at the end of the subsection 5.1.

% end of subsect 5.2.

\bigskip
\bigskip

Let us now rewrite the effective action in the form
 $$
\Gamma_{(1)} = -{1\over 2}\zeta'_{tot}(0)
\eqno(5.110)
$$
where
 $$
 \zeta_{tot}(p) = \zeta_{\Delta}(p) - 2\zeta_F(p) - \zeta_{K}(p) +
 \zeta_{N}(p)
\eqno(5.111)
 $$
 is the total  $\zeta$-function.

Collecting all obtained expressions for the heat kernels it is found
for the total $\zeta-$ function
$$
\eqalignno{
\zeta_{tot}(p) &= \int dx (4\pi)^{-d/2}{\m^{2p}\over \Gamma(p)}
\int\limits_0^\infty
d t\,t^{p-d/2-1}\Biggl\{r(d-2) &\cr
&+ 2\sum_{\a>0}
\prod_{1\le i\le q}\left({tH_i(\a)\over \sinh(tH_i(\a))}\right)
\left(d-2 + 4\sum_{1\le j\le q}\sinh^2(tH_j(\a))\right)&\cr
&-2^{[d/2]}\exp(-tM^2_{(0)})\left(R + \sum_{\n\ne 0} d_\n\exp(-i t M^2(\n))
\prod_{1\le i \le q}\left(tH_i{(\n)}
\coth\left(tH_i{(\n)}\right)\right)\right)&\cr
&+\exp(-tQ_{(0)})\left(\tilde R + \sum_{\tilde\n\ne 0}
d_{\tilde\n}\exp(-i t Q{\tilde\n})
\prod_{1\le i\le q}\left({tH_i{(\tilde\n)}\over
\sinh\left(tH_i{(\tilde\n)}\right)}\right)\right)\Biggr\}&
(5.112)\cr}
 $$
where $\a$ are the roots of the algebra, $\n$ and  $\tilde\n$ are the
weights and $R$ and $\tilde R$ are the numbers of zero weigths of the
representations realized by the spinor and scalar fields respectively,
$H_i(\a)$, $H_i(\n)$ and $H_i{(\tilde\n)}$ are the invariants of the tensors
$F_{\m\n}^a\a_a$, $F^a_{\m\n}\n_a$ and $F^a_{\m\n}\tilde\n_a$ respectively,
defined from the equations of the form (5.49) or (5.49a).

After taking a concrete gauge group and the matter field representations one
can obtain from here more explicit expressions for the effective action.

\bigskip
\bigskip
\centerline{\bf YI. CONCLUDING REMARKS}
\bigskip

The present paper is, as a matter of fact,  a further development of our
papers
[1-7].
Here we have discussed some ideas connected with the point that was left
aside in previous papers, namely, the problem of calculating the low-energy
limit of the effective action in quantum gravity and gauge theories and have
presented an overview of recent progress on this subject [67, 68].

We have analyzed in detail  the status of the low-energy limit in quantum
gauge theories and quantum gravity and stressed the central role
playing in low-energy calculations by the Lie group (3.65)-(3.67) that
naturally appears when generalizing consistently the low-energy limit to
quantum gravity.

We have proposed a number of promising, to our mind, approaches for
calculation of the low-energy effective action and have calculated explicitly
the low-energy effective action in flat space. In particular, a Yang-Mills
model with arbitrary matter fields is considered and the problem of the
stability of Savvidy-type vacuum structure
[65, 66]
with constant chromomagmetic fields is
analyzed.  It is shown that such a vacuum state can be stable only in the
case when more than one constant chromomagnetic fields are present and the
values of these  fields differ not greatly from each other. As a
consequence we concluded that this is possible only in spaces with
dimensions not less than five    $d\geq 5$. This fact can be useful in
investigating the QCD condensates etc..

Thereby we realized the idea of partial summation of the terms without
derivatives in local Schwinger - De Witt expansion for computing the
effective action that was suggested in
[13,5,6].

Of course, there are left many unsolved problems. We still do not know how to
calculate the low-energy effective action in general case of covariantly
constant curvatures (3.12). Besides, it is not perfectly clear how to do the
analytical continuation of Euclidean  low-energy effective action to the
space of Loretzian signature for obtaining physical results.  The methods of
the proof of the possibility of such a continuation
[48]
are based essentially on the expansion in curvatures, i.e. are
valid, strictly speaking, in weak background fields and trivial topology.

\bigskip
\centerline{\bf ACKNOWLEDGEMENTS}
\medskip
I would like to thank G. A. Vilkovisky for many helpful discussions
and R. Schimming and J. Eichhorn for their hospitality at the University
of Greifswald.
\vfill
\eject

%
% References
%

\centerline{\bf REFERENCES}
\bigskip
\item{[1]} I. G. Avramidi, Teor. Mat. Fiz. {\bf 79},  219 (1989)
\item{[2]} I. G. Avramidi, Yad. Fiz. {\bf 49}, 1185 (1989)
\item{[3]} I. G. Avramidi, Phys. Lett. {\bf B236}, 443 (1990)
\item{[4]} I. G. Avramidi,  Phys. Lett. {\bf B238}, 92 (1990)
\item{[5]} I. G. Avramidi,  {\it The covariant technique for calculation of
            one-loop effective action}, Preprint University of Karlsruhe, \
            KA - THEP - 2 - 1990, 33 p., (1990)
\item{[6]} I. G. Avramidi,  Nucl. Phys. {\bf B355},712 (1991)
\item{[7]} I. G. Avramidi, Yad. Fiz. {\bf 56},245 (1993)
\item{[8]} J. S. Schwinger, Phys. Rev.{\bf 82}, 664 (1951)
\item{[9]} B. S. De Witt, {\it Dynamical Theory of Groups and Fields}
           (Gordon \& Breach, New York, 1965)
\item{[10]} B. S. De Witt, Phys. Rep. {\bf C19}, 296 (1975)
\item{[11]} B. S. De Witt, in {\it General Relativity}, edited by S. Hawking
		and W. Israel (Cambridge University Press, Cambridge, 1979)
\item{[12]} B. S. De Witt, in {\it Relativity, Groups and Topology II},
edited by B. S.  De Witt and R. Stora (North Holland, Amsterdam, 1984) p. 393
\item{[13]} G. A. Vilkovisky, in {\it Quantum Theory of Gravity}, edited by S.
	     Christensen (Hilger, Bristol, 1984) p. 169
\item{[14]} E. S. Fradkin and A. A. Tseytlin, Nucl. Phys. {\bf B234}, 472
            (1984)
\item{[15]} G. A. Vilkovisky, Class. Quant. Grav. {\bf 9}, 895 (1992)
\item{[16]} I. L. Buchbinder and S. D. Odintsov, Fortschr. Phys. {\bf 37},
		225  (1989)
\item{[17]} I. L. Buchbinder, S. D. Odintsov, I. L. Shapiro, Riv. Nuovo Cim.
             {\bf 12}, 1 (1989)
\item{[18]} S. D. Odintsov, Fortschr. Phys. {\bf 38}, 371 (1990)
\item{[19]} J. A. Zuck, Phys. Rev.  {\bf D33}, 3645 (1986)
\item{[20]} N.  D.  Birrell and P. C. W. Davies, {\it Quantum Fields  in
	    Curved Space} (Cambridge University Press, Cambridge, 1982)
\item{[21]} A. O. Barvinsky and G. A. Vilkovisky, Phys. Rep. { \bf C119}, 1
	     (1985)
\item{[22]} G. A. Vilkovisky, {\it Heat Kernel: Recontre Entre Physiciens et
	    Mathematiciens}, Preprint CERN-TH.6392/92 (1992), to be published
	    in Proceedings of Strasbourg Meeting between Physicists and
	    Mathematicians (1991), Inst. de Recherche Math.  Avancee, Univ.
	    Louis Pasteur, Strasbourg
\item{[23]}  J. Hadamard, {\it Lectures on Cauchy's Problem},
		in {\it Linear Partial Differential Equations}
		(Yale University Press, New Haven, 1923)
\item{[24]}  S. Minakshisundaram and A. Pleijel, Can. J. Math.
 		{\bf 1}, 242 (1949)
\item{[25]}  H. P. McKean and I. M. Singer, J Diff. Geom. {\bf 1}, 43 (1967)
\item{[26]} R. T. Seely, Proc. Symp. Pure Math. {\bf 10}, 288 (1967)
\item{[27]}  M. Atiyah, R. Bott and V. K. Patodi, Invent. Math.
		{\bf 19}, 279 (1973)
\item{[28]} P. B. Gilkey, J. Diff. Geom. {\bf 10}, 601 (1975)
\item{[29]} P. B. Gilkey, Proc. Symp. Pure Math. {\bf 27}, 265 (1975)
\item{[30]} P. B. Gilkey, Compositio Math. {\bf 38}, 201 (1979)
\item{[31]} P. B. Gilkey, {\it Invariance Theory, the Heat Equation and
	    the Atiyah - Singer Index Theorem}
	    (Publish or Perish, Wilmington, Delaware (USA), 1984)
\item{[32]} R. Schimming, Beitr. Anal. {\bf 15}, 77 (1981)
\item{[33]} R. Schimming, Math. Nachr. {\bf 148}, 145 (1990)
\item{[34]} R. Schimming, {\it Calculation of the Heat Kernel Coefficients,}
	    in B. Riemann Memorial Volume, edited by T. M. Rassias,
	    (World Scientific, Singapore, to be published)
\item{[35]} S. A. Fulling, SIAM J. Math. Anal. {\bf 13} (1982) 891
\item{[36]} S. A. Fulling and G. Kennedy, in {\it The Physics of Phase Space},
            (Lecture  Notes in Physics {\bf 278}), edited by Y. S.  Kim and
	    W. W. Zachary, (Springer, Berlin, 1986), p.  407
\item{[37]} S. A. Fulling and G. Kennedy, in {\it Differential Equations and
            Mathematical Physics,} (Lecture Notes in Mathematics {\bf 1285}),
      edited by I. W. Knowles and Y.  Saito (Springer, Berlin, 1987), p. 126
\item{[38]} S. A. Fulling and G. Kennedy, Transact. Amer. Math. Soc.  {\bf
            310}, 583 (1988)
\item{[39]} S. A. Fulling, J. Symbolic Comput. {\bf 9}, 73 (1990)
\item{[40]} F. H.  Molzahn, T.  A.  Osborn and S. A. Fulling, Ann. Phys. (USA)
	    {\bf 204}, 64 (1990)
\item{[41]} H. Widom, Bull. Sci. Math. {\bf 104}, 19 (1980)
\item{[42]} V. P. Gusynin, Phys. Lett. {\bf B255}, 233 (1989)
\item{[43]} V. P. Gusynin, Nucl. Phys. {\bf B333}, 296 (1990)
\item{[44]} V. P. Gusynin and V. A. Kushnir, Class. Quant. Grav. {\bf 8}, 279
	       (1991)
\item{[45]} P. Amsterdamski, A. L. Berkin and D. J. O'Connor, Class. Quant.
		Grav.  {\bf 6}, 1981 (1989)
\item{[46]} A. E. M. Van de Ven, Nucl. Phys. {\bf B250}, 593 (1985)
\item{[47]} V. P. Frolov and A. I. Zelnikov, Phys. Rev. {\bf D35}, 3031 (1987)
\item{[48]} A. O. Barvinsky and G. A. Vilkovisky, Nucl. Phys.
		{\bf B282}, 163 (1987)
\item{[49]} A. O. Barvinsky and G. A. Vilkovisky, Nucl. Phys.
		{\bf B333}, 471 (1990)
\item{[50]} A. A. Ostrovsky and G. A. Vilkovisky, J. Math. Phys.
		{\bf 29}, 702 (1988)
\item{[51]} A. O. Barvinsky, Yu. V. Gusev, V. V. Zhytnikov and G. A.
		Vilkovisky, {\it Covariant Perturbation Theory (IY),}
		Preprint University of Winnipeg (1993)
\item{[52]} G. A. Vilkovisky, {\it Quantum Theory of Gravitational
		Collapse}, Lectures at the University of Texas at Austin
		(1989); (unpublished)

\item{[53]} B. S. De Witt and R. W. Brehme, Ann. Physics {\bf 9}, 220 (1960)

\item{[54]} T. P.  Branson  and P.  B.  Gilkey, Comm. Part. Diff. Eq.
		{\bf 15}, 245 (1990)
\item{[55]} M. van den Berg and P. B. Gilkey, {\it Heat content asymptotics
		of a Riemannian manifold with boundary}, Preprint University
		of Oregon (1992) (to be published)
\item{[56]} S. Desjardins and P. B. Gilkey, {\it Heat  content asymptotics
		for operators of Laplace type with Newmann boundary
		conditions}, Preprint (1992) (to be published)
\item{[57]} M.van den Berg, S. Desjardins and P. B. Gilkey, {\it
		Functorality and heat content asymptotics for operators of
		Laplace type}, Preprint (1992)
\item{[58]} D. M. Mc Avity and H. Osborn, Class. Quant. Grav. {\bf 8},
		603 (1991)
\item{[59]} D. M. Mc Avity, Class. Quant. Grav. {\bf 9}, 1983 (1992)
\item{[60]} T. Branson, P. B. Gilkey and B. \O rsted, Proc. Amer.  Math.
	    Soc.  {\bf 109}, 437 (1990)
\item{[61]} B. F. Dubrovin, S. P.  Novikov and A. T. Fomenko, {\it The
	    Modern Geometry: Methods and Applications} (Nauka, Moscow, 1979)
\item{[62]} J. L. Synge, {\it Relativity. The General Theory}
	    (North-Holland, Amsterdam, 1960)
\item{[63]} B. S. De Witt, in  {\it Quantum Field Theory and Quantum
	    Statistics}, vol. 1, edited by I. A.  Batalin, C.  J. Isham and
	    G.  A.  Vilkovisky (Hilger, Bristol, 1987) p. 191
\item{[64]} A. O.  Barut and R. Raczka, {\it Theory of Group
	Representations and Applications}, (PWN-Polish Sci. Publ.,
		Warszawa, 1977)
\item{[65]} G. K. Savvidy, Phys. Lett.  {\bf B71}, 133 (1977)
\item{[66]} N. K. Nielsen and P. Olesen, Nucl. Phys. {\bf B144}, 376 (1978)
\item{[67]} I. G. Avramidi,  Phys. Lett. {\bf B305}, 27 (1993)
\item{[68]} I. G. Avramidi,  Covariant algebraic calculation of the one-loop
effective potential in non-Abelian gauge theory and a new approach to
stability problem, Preprint University Greifswald (1994), (to be published)

% end of references

\vfill
\eject
\bye